\documentclass[a4paper,fleqn,usenatbib,useAMS,usedcolumns]{mnras}
\usepackage{newtxtext,newtxmath}

\usepackage[T1]{fontenc}
\usepackage{ae,aecompl}
\usepackage{comment}
\usepackage{threeparttable}
\usepackage{booktabs, caption, makecell}

\usepackage{graphicx}	
\usepackage{amsmath}	
\usepackage{gensymb}
\usepackage{siunitx}
\usepackage{comment}

\newcommand{\hi}{\mbox{H\,{\sc i}}}

\newcommand{\civ}{\mbox{C\,{\sc iv}}}

\newcommand{\nv}{\mbox{N\,{\sc v}}}

\newcommand{\siiv}{\mbox{Si\,{\sc iv}}}
\newcommand{\siv}{\mbox{S\,{\sc iv}}}

\newcommand{\RN}[1]{%
  \textup{\uppercase\expandafter{\romannumeral#1}}%
}

\def\h2{$\rm H_2$}
\def\Nh2{$N$(H${_2}$)}

\def\lya{\ensuremath{{\rm Ly}\alpha}}

\def\kms{km\,s$^{-1}$}

\def\zabs{$z_{\rm abs}$}
\def\zem{$z_{\rm em}$}

\def\21{21-cm}

\def\t0{T$_{0}$}

\def\c21{$C_{21}$}

\def\J13{J$1322+0524$}

\title[BALs in J1322+0524]{ Coordinated time variability of multi-phase ultra-fast outflows in J132216.25+052446.3 \thanks{Based on observations collected at Southern African Large Telescope (SALT; Programme IDs 2015-1-SCI-005, 2015-2-SCI-026, 2018-1-SCI-009, 2019-1-SCI-019, 2019-2-SCI-016, 2020-2-SCI-014, 2021-1-SCI-005).}}

\author[Aromal et al.]{
P. Aromal$^{1}$\thanks{E-mail: aromal@iucaa.in (PA)},
R. Srianand$^{1}$,
and P. Petitjean$^{2}$
\\
$^{1}$IUCAA, Postbag 4, Ganeshkind, Pune 411007, India\\
$^{2}$ Institut d'Astrophysique de Paris, Sorbonne Universit\'e and CNRS, 98bis boulevard Arago, 75014 Paris, France\\
}

\date{Accepted XXX. Received YYY; in original form ZZZ}

\pubyear{2015}

\begin{document}
\label{firstpage}
\pagerange{\pageref{firstpage}--\pageref{lastpage}}
\maketitle

\begin{abstract}
We present a time variability  analysis of broad absorption lines (BAL; spread over the velocity range of 5800-29000 \kms) seen in the spectrum of J132216.25+052446.3 (\zem = 2.04806) at ten different epochs spanning over 19 years. 
The strongest absorption component (BAL-A; spread over 5800-9900 \kms) is made up of several narrow components having velocity separations close to \civ\ doublet splitting. The \civ, \nv\ and \siiv\ absorption from BAL-A  show correlated  optical depth variability without major changes in the velocity structure. 
A very broad and shallow absorption (BAL-C;  spread over the velocity range 15000-29000 \kms) emerged during our monitoring period coinciding with a dimming episode of \J13.
All the identified absorption lines show correlated variability with the equivalent widths increasing with decreasing flux.
This together with the \civ\ emission line variability is consistent with ionization being the main driver of the correlated variability.
The observed UV-continuum variations are weaker than what is required by the photo-ionization models.
This together with a scatter in the \civ\ equivalent width at a given continuum flux can be understood if  variations of the \civ\ ionizing photons are much larger than that of the UV continuum,  the variations in the ionizing photon and UV fluxes are 
not correlated and/or the covering factor of the flow varies
continuously.
We suggest
BAL-A is produced by a stable clumpy outflow located beyond the broad emission line region
and  BAL-C is a newly formed wind component located near the accretion disk  and both respond to changes in the ionizing continuum.
\end{abstract}




\section{Introduction}

Broad Absorption Line (BAL) quasars are defined by the presence of absorption lines with large velocity widths ($\sim$ few 1000 \kms) and ejection velocities (reaching up to few 10,000 \kms) \citep{Weymann1991}.
It is believed that these outflows could play an
important role in 
the central black hole growth, the host galaxy evolution \citep{ostriker2010,kormendy2013} and the chemical enrichment of the intergalactic medium (IGM). 
Detailed investigation of the time variability of BAL profiles is important to obtain
tight constraints on the BAL lifetime, location of the outflow etc., and provide significant insights on the origin and physical mechanisms driving the flow. 
BAL variability includes extreme optical depth variations like emergence, disappearance and kinematic shift of BALs \citep{Filiz2013,mcgraw2017,rogerson2018,vivek2018,cicco2018}. Possible origins of BAL variability include : (i) large variations in quasar ionizing flux, (ii) changes in covering factor ($f_c$) of the outflow with respect to the central source, (iii) transverse motion of the outflow perpendicular to our line of sight. 

In recent years, using the Southern African Large Telescope (SALT) we started to investigate the time variability of a sample of 62 quasars from SDSS DR15 \citep{paris2017} which show BALs at outflow velocities greater than 15,000 km~s$^{-1}$ in their spectra (corresponding to Ultra Fast Outflows, UFOs). 
This monitoring programme has revealed several interesting BAL quasars. In \citet{aromal2021} we presented a detailed analysis of J162122.54+075808.4, that shows emergence and acceleration signatures of new BAL components at large velocity. 
In this paper, we present a detailed analysis of another interesting BAL QSO (J132216.25+052446.3, hereafter denoted as J1322+0524) for the following reasons: (i)
Presence of a broad absorption component consisting of multiple narrow absorption features that are separated roughly by C~{\sc iv} doublet splitting. Such a signature (usually referred to as "line-locking") is considered as an evidence for line-driven acceleration in disk wind models. 
(ii) Emergence of a very high velocity flow (with ejection velocity in the range 15000-28000 \kms) revealed by C~{\sc iv} and N~{\sc v} absorption. Absorption with such large velocities and spread are expected 
in standard disk-wind models. Thus this component could represent a newly ejected material from the accretion disk in contrast to transverse motion of clouds usually invoked to understand newly emerged absorption.
(iii)  Availability of ten epochs of spectroscopic monitoring data spreading over 19 years and good photometric coverage in ZTF over the last 5 epochs (with at least one spectrum per year)  provides a possibility to constrain different time-scales related to this BAL quasar. (iv) Presence of an apparent correlated variability between different broad absorption components and \civ\ broad emission line.

The paper is organized as follows. In section~\ref{sec:obs}, we present the details of the spectroscopic observations of \J13\ using SALT and data reduction. 
In section~\ref{sec:abs}, we provide the details of the \civ\ BAL components and quantify their rest equivalent width and kinematic variability. We probe the possible correlations in the variability of the \civ\ rest equivalent widths of the different BAL components and the \civ\ broad emission line. In section~\ref{sec:continuum}, we study the long-term rest UV continuum flux  variability of \J13\ using all the available spectra and photometric light-curves.
We also study the possible correlation between the \civ\ rest equivalent width and the rest frame near-UV (NUV) continuum variabilities.
In section~\ref{sec:discuss}, we discuss our results in the framework of simple photo-ionization and disk-wind model predictions.
We summarize  our main findings in section~\ref{sec:summary}.

\section{Observations $\&$  Data used in this study}
\label{sec:obs}
\begin{table*}
\begin{threeparttable}
    \centering
\caption{Log of observations and details of spectra obtained at different epochs 
}
 \begin{tabular}{cccccccccccc}
  \hline
  Epoch & Telescope & \multicolumn{2}{c}{Date of observations} & Exposure  & Spectral  & S/N \tnote{a} & $W_{\text{\civ}}^A$ \tnote{b} & $W_{\text{\civ}}^B$ \tnote{c} & $W_{\text{\civ}}^C$ \tnote{d} \\ 
&       used  & (M/D/Y) & (MJD) &  time (s)& res. (\kms) & &(\AA) & (\AA) & (\AA)\\
  \hline\hline
 1& SDSS & 04-12-2002 & 52376  & 7207 & 150 & 13.37 & $11.34 \pm 0.16$ & $1.88 \pm 0.17$ & ....\\
 2& SDSS & 03-13-2011 & 55633  & 7207 & 150 & 26.41 & $8.57 \pm 0.08$ & $0.73 \pm 0.08$ & .... \\
 3& SDSS & 05-22-2011 & 55703  & 7207 & 150 & 26.08 & $8.25 \pm 0.08$ & $0.72 \pm 0.09$ & ....\\
 4& SALT & 06-21-2015 & 57194 & 2400 & 304 & 29.37 & $3.57 \pm 0.07$ & $0.73 \pm 0.07$  & $ 6.00 \pm 0.13$\\
 5& SALT & 04-15-2016 & 57493  & 2400 & 304 & 22.80 & $5.06 \pm 0.11$ & $1.17 \pm 0.09$ & $ 8.56 \pm 0.17$ \\
 6& SALT & 05-07-2018 & 58245  & 2074 & 304 & 15.62 & $9.75 \pm 0.16$ & $1.97 \pm 0.14$ & $ 13.35 \pm 0.26$\\
 7& SALT & 06-09-2019 & 58643  & 2400 & 304 & 19.89 & $11.66 \pm 0.11$ & $2.54 \pm 0.11$ & $12.80 \pm 0.20$\\    
 8& SALT & 03-23-2020 & 58931  & 2400 & 304 & 35.55 & $7.34 \pm 0.06$ & $1.02 \pm 0.06$ & $6.33 \pm 0.11$\\    
 9 & SALT & 02-17-2021 & 59262  & 2500 & 304 & 42.46 & $9.13 \pm 0.05$ & $1.67 \pm 0.05$ & $8.04 \pm 0.09$\\    
 10& SALT & 05-10-2021 & 59344  & 2500 & 304 & 54.69 & $6.49 \pm 0.04$ & $0.92 \pm 0.04$ & $5.80 \pm 0.07$\\    
  \hline
 \end{tabular}
 \begin{tablenotes}\footnotesize
    \item[a] Signal-to-noise ratio per-pixel calculated over the wavelength range 4900 - 5200 \AA~.
    \item[b,c,d] Total \civ\ rest equivalent width of A, B and C components obtained by integrating over distinct regions in the absorption profile respectively (see Fig.~\ref{fig:abs_marked_fig}).

 \end{tablenotes}
\label{tab_obs}
\end{threeparttable}
\end{table*}

\begin{table}
\begin{threeparttable}
    \centering
\caption{Equivalent width variations of BAL A component
}
 \begin{tabular}{ccccccccccc}
  \hline
   Epoch & \multicolumn{4}{c}{Rest equivalent width (\AA) of}  \\ 
         & \civ\ & \siiv\ & \nv\ & \lya\ \\
  \hline\hline
  1 &  $11.34 \pm 0.16$ & $3.37 \pm 0.25$ & - & - \\
  2 &  $8.57 \pm 0.08$ & $\le$0.36 & $8.72 \pm 0.12$ & $4.06 \pm 0.25$ \\
  3 &  $8.25 \pm 0.08$ & $\le$0.39 & $7.74 \pm 0.13$ & $5.41 \pm 0.21$ \\
  4 &  $3.57 \pm 0.07$ & $\le$0.27 & $5.29 \pm 0.06$ & $3.80 \pm 0.09$ \\
  5 &  $5.06 \pm 0.11$ & $\le$0.36 & $6.52 \pm 0.07$ & $4.16 \pm 0.11$\\
  6 &  $9.75 \pm 0.16$ & $3.13 \pm 0.18$ & $9.20 \pm 0.11$ & $5.30 \pm 0.20$ \\
  7 &  $11.66 \pm 0.11$ & $2.82 \pm 0.14$ & $10.03 \pm 0.09$ & $4.87 \pm 0.14$ \\
  8 &  $7.34 \pm 0.06$ & $\le$0.24 & $7.70 \pm 0.05$ & $3.94 \pm 0.09$ \\
  9 &  $9.13 \pm 0.05$ & $2.46 \pm 0.06$ & $8.99 \pm 0.04$ & $4.32 \pm 0.07$ \\
  10 & $6.49 \pm 0.04$ & $\le$0.15 & $7.31 \pm 0.04$ & $4.11 \pm 0.06$ \\
  \hline
 \end{tabular}

\label{tab_bal_a}
\end{threeparttable}
\end{table}


Spectra of \J13 were obtained during three epochs in the SDSS/BOSS  (Sloan Digital Sky Survey/ Baryon Oscillation Spectroscopic Survey)  survey.  For more than six years since the year 2015, we have been carrying out 
spectroscopic monitoring of \J13 using the SALT \citep[][]{buckley2005}.
Details of all the available spectra are summarized in Table~\ref{tab_obs}. We used the Robert Stobie Spectrograph \citep[RSS,][]{Burgh2003,Kobulnicky2003} at SALT in the long-slit mode using a 1.5" wide slit and the PG0900 grating (kept at a position angle, PA = 0.0 for epochs 4, 5 and  7, PA = 85.0 for epoch 6 and PA = 80.0 for epochs 8, 9 and 10).  With GR angle=$12.5^{\circ}$ and CAM angle=$25.0^{\circ}$, this setting provides a wavelength coverage of $3210-6300$ {\AA} excluding the 4216-4273 {\AA} and 5276-5331 {\AA} regions falling in the CCD gaps. 

The raw CCD frames have been preliminary processed using the SALT data reduction pipeline \citep{Crawford2010}. We used the standard {\sc iraf}\footnote{{\sc iraf} is distributed by the National Optical Astronomy Observatories, which are operated by the Association of Universities for Research in Astronomy, Inc., under cooperative agreement with the National Science Foundation.} procedures to reduce the resulting 2D spectra. Flat-field corrections and cosmic ray zapping were applied to all science frames. We extracted the one dimensional quasar spectrum from the background subtracted 2D science frames from each epoch using the {\sc iraf} task ``apall". Wavelength calibration was performed using standard Argon lamp spectra. 
In addition, skylines from the wavelength calibrated spectrum were matched with the sky line atlas provided by SALT and, if needed, corrections were applied to increase wavelength accuracy. Similarly, flux calibration was performed using reference stars (G93-48 and LTT4364) observed close to our observing nights. 

The spectral resolution and average signal-to-noise ratio (S/N) obtained are summarized in columns 6 and 7 of Table~\ref{tab_obs}. Our SALT spectra typically have a spectral resolution of $R \sim 830$ (roughly a factor of two smaller than that of SDSS) at central wavelength $4780$ \AA\ and S/N in the range 13-55  per-pixel. In total, we have spectra of \J13 obtained during 10 epochs spreading over 19 years. We denote them by epoch-1, epoch-2, etc., in chronological order (given in column 1 of Table~\ref{tab_obs}).
In the SDSS catalog \citep{paris2012} the quoted emission redshift of \J13 is \zem = $2.0574 \pm 0.0002	$. \citet{Hewett2010} derive a systemic redshift of \zem = $2.0498 \pm 0.0006$ from the fit of the C~{\sc iii}] emission. We use this latter value as the systemic redshift for all discussions in this paper. Note that given the high velocities of the BALs we study here, the exact emission redshift has  little influence on our study.

\begin{figure*}
    \centering
    \includegraphics[viewport=150 40 1560 890, width=\textwidth,clip=true]{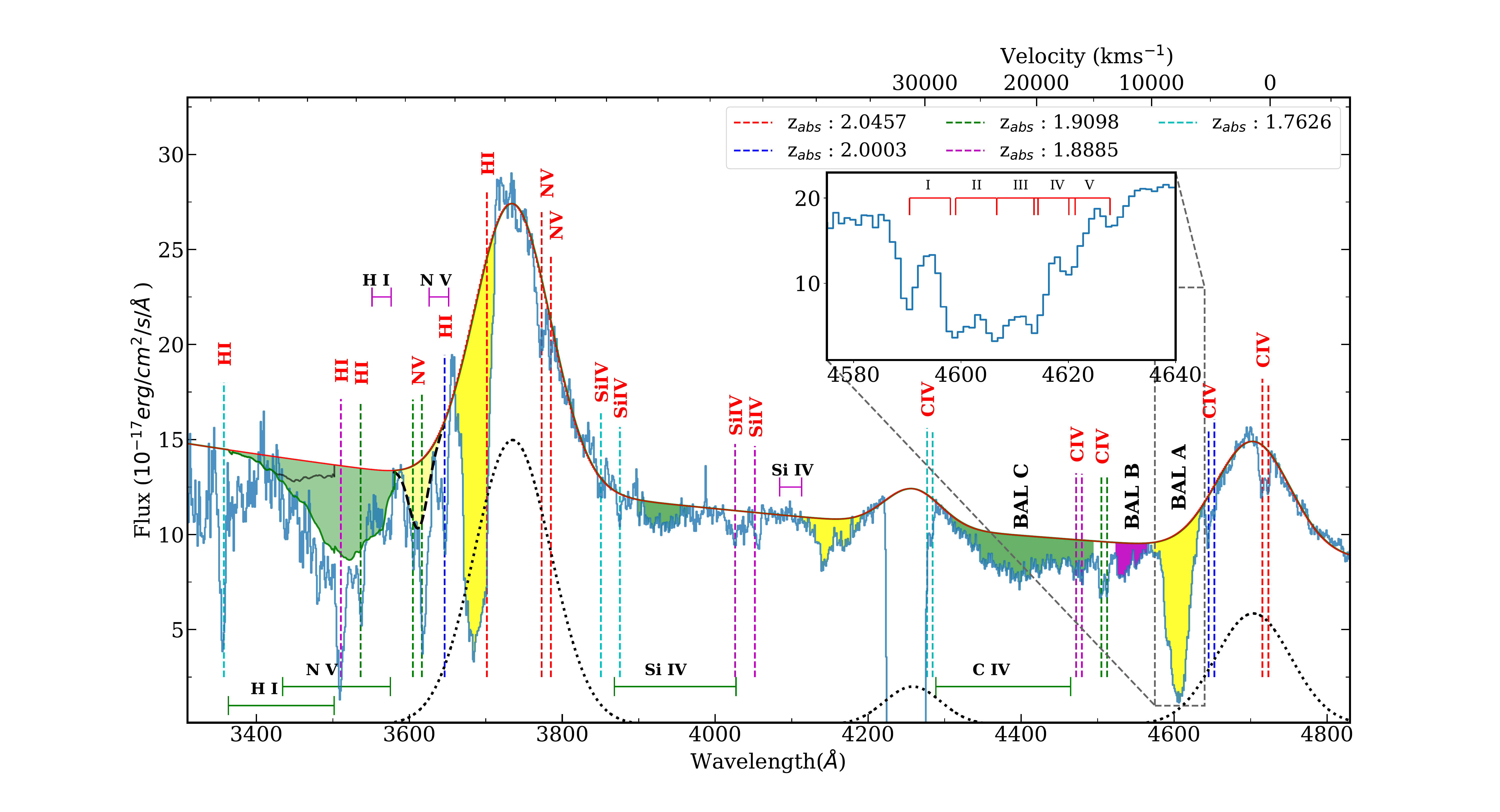}
    \caption{
    SALT spectrum of \J13\ obtained during MJD 59262 (epoch-9) overlaid with the best fitted continuum (in red). The emission line contributions are shown using dotted Gaussians. We mark different absorption lines (identified with labels) from 5 narrow absorption systems  identified based on \civ\ doublets (different colors). For easy discussions, we segregate the \civ\ broad absorption profile into three components (i.e BAL-A, B and C) shown with shaded regions having  different colours. Inset shows the multiple component nature of BAL-A with possible signatures of line-locking at \civ\ velocity splitting in epoch-2. The velocity scale for \civ\ BAL absorption with respect to the systemic redshift (\zem = 2.0498) is provided at the top. The expected wavelength range for different absorption lines from BAL-B and BAL-C are indicated with magenta and green horizontal lines respectively. In the case of \nv\ and \lya\ (both affected by \lya\ contamination) the shaded region has been scaled from the \civ\ profile for illustration. 
    }
    \label{fig:abs_marked_fig}
\end{figure*}

\begin{figure*}
    \centering
    \includegraphics[viewport=80 40 1230 1250, width=\textwidth,clip=true]{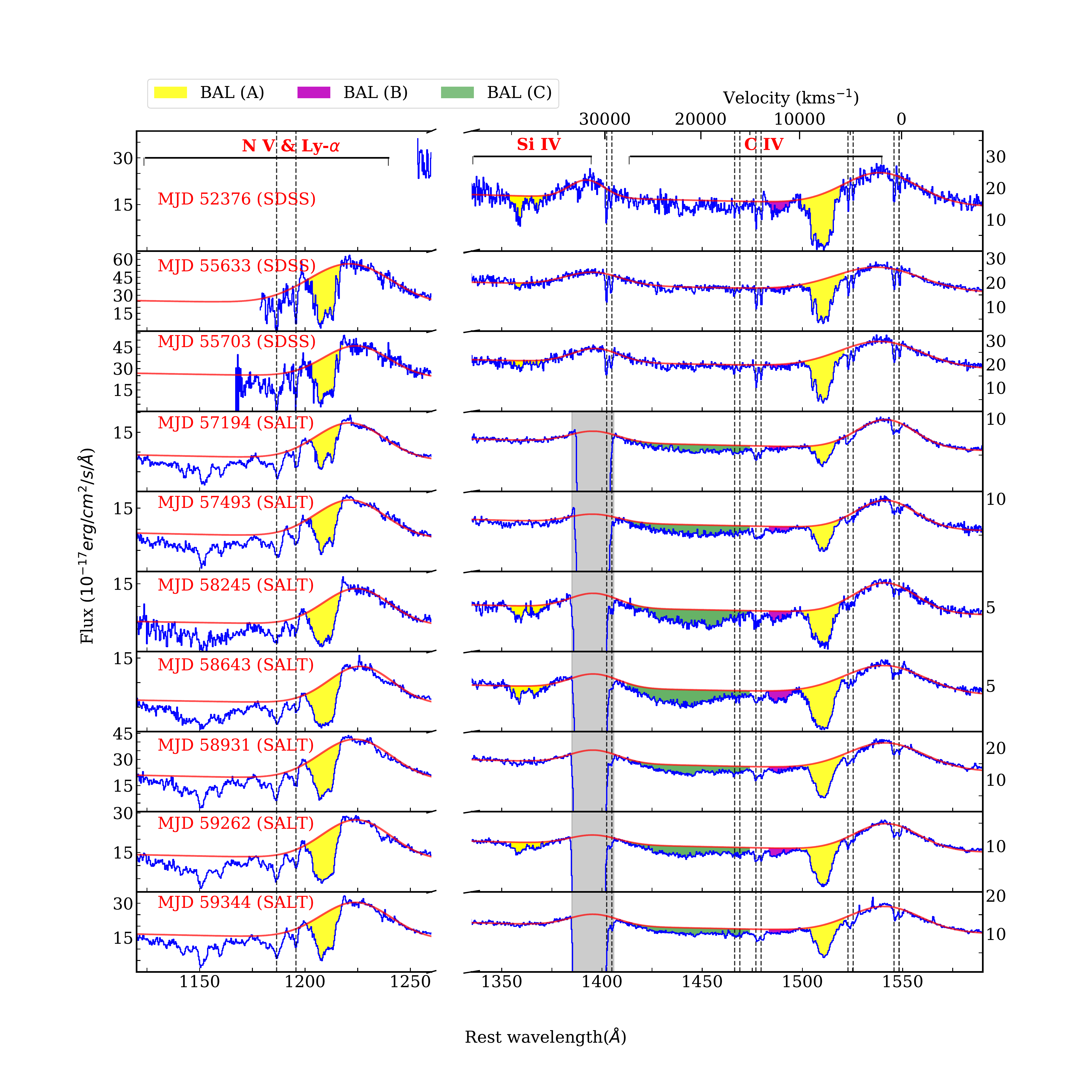}
    \caption{
    Continuum fits (red) to the rest frame spectra (blue) of \J13\ defined with respect to \zem = 2.0498. Different narrow absorption lines are marked with vertical dashed lines. 
    Absorption line variability of broad components identified with shaded regions are apparent. 
    The velocity scale for \civ\ BAL absorption with respect to the systemic redshift (\zem = 2.0498) is provided at the top. Gray shaded regions represent the CCD gaps in SALT spectra.
    }
    \label{fig:norm_spectra}
\end{figure*}

Continuum fitting is very important in the study of BALs as the measurement of equivalent width of various lines is highly dependent on its accuracy. For \J13\, we use the conventional power law + Gaussians fit, where a power-law is used to fit the line-free regions (excluding the wavelength range with narrow and broad absorption and emission lines (BEL)) and single Gaussians are used to fit \civ, \siiv\ and \nv\ BELs. For the power-law fit we considered the following line-free regions in the rest frame of the quasar : 1335.0$-$1350.0 \AA, 1635.0$-$1730.0 \AA, 1755.0$-$1820.0 \AA, 1970.0$-$2700.0 \AA. 
In Fig.~\ref{fig:abs_marked_fig}, we show the SALT spectrum of \J13 obtained during epoch-9 (MJD 59262) together with the best fitted continuum (red). 
Various prominent absorption lines are also identified in this figure. We provide the details of these systems and their variability nature in the following sections. Continuum fits to spectra obtained in the different epochs are shown in Fig~\ref{fig:norm_spectra}.

As in \citet{aromal2021}, we correlate the absorption line variability with the continuum variability using available broad band photometric light curves. For this, we obtained publicly available photometric light-curves of \J13 from the Panoramic Survey Telescope and Rapid Response System \citep[Pan-STARRS;][]{panstarrs2016}, the Palomar Transient Factory \citep[PTF;][]{Law2009} and the Zwicky Transient Facility \citep[ZTF;][]{zptf2019b,zptf2019a} surveys. Pan-STARRS provides photometric data of \J13 for five broad band filters, i.e., g, r, i, z and y whereas ZTF gives the same for the g, r and i bands. Details of photometric variability of \J13\ are discussed in section~\ref{sec:continuum}.

\section{Analysis of Absorption features}
\label{sec:abs}

In this section, we mainly focus on the different absorption lines detected in the spectra of \J13 and their time variability.

\subsection{Narrow intervening absorption lines}

We identify five \civ\ narrow absorption line systems in the spectra of \J13 in all epochs. 
In Fig~\ref{fig:abs_marked_fig}, we identify these systems and mark different ionic absorptions like \civ, \siiv, \nv\ and \lya\ originating from these systems. The redshifts of these systems are \zabs = 1.7626, 1.8885, 1.9098, 2.0003 and 2.0457. The last two systems are within 5000 \kms\ to the quasar's systemic redshift and therefore satisfy one of the definitions of associated absorption  \citep{foltz1986}. %
We do not detect any statistically significant change in the rest equivalent width of \civ\ or any other absorption from both these systems. As the spectra used here are of low resolution, we are not in a position to measure the covering factor in these cases. Presence of partial coverage would have confirmed these systems to be physically associated with \J13\
\citep{Hamann1997, Barlow1997}.

The first three systems show strong  associated \lya\ absorption.
The \zabs = 1.8888 system also shows absorption from low-ionization species (Mg~{\sc ii}, Fe~{\sc ii}, Si~{\sc ii} and Al~{\sc ii}) as well. We also detect weak Mg~{\sc ii} absorption associated with the
\zabs = 1.7626 system in the SDSS spectrum. In addition to the above mentioned 5 \civ\ systems, we detect an intervening system at \zabs = 1.6050 based on the Mg~{\sc ii} doublet. We also detect Fe~{\sc ii} and Al~{\sc ii} absorption lines from this system but no \civ\ absorption.
We do not observe any significant variability in the rest equivalent widths of  \civ\ and/or other absorption lines from these 6  narrow absorption systems between our spectroscopic epochs. 
We used these narrow absorption lines 
to check our wavelength calibration accuracy.

\subsection {Broad absorption line}

\J13 was classified as a BAL quasar  based on the presence of a broad \civ\ absorption at \zabs$\sim$1.975 in the first SDSS spectrum (epoch-1). The \civ\ absorption covers an ejection velocity range of 5800-13300 \kms\ with respect to the systemic redshift and is split in two sub-components.
Absorption from Si~{\sc iv}, N~{\sc v} and \lya\ are clearly detected (see yellow and magenta shaded regions in Fig~\ref{fig:abs_marked_fig}).  We do not detect Al~{\sc iii} absorption in any of our spectrum. The wavelength range of Mg~{\sc ii} absorption is covered only in SDSS spectra and we do not detect Mg~{\sc ii} absorption.
We refer to these systems as BAL-A and BAL-B in our discussions below. In our subsequent monitoring using SALT observations, we identify an additional newly emerged broad \civ\ absorption spread up to an ejection velocity of $\sim$29000 \kms (see green shaded regions in Figs~\ref{fig:abs_marked_fig} and~\ref{fig:norm_spectra}). 
For the ease of discussions, we name this emerging BAL as
BAL-C. 
Below we provide detailed description of these three BAL components.

\subsubsection{Observed properties of BAL-A :}

\J13\ was identified as a BAL quasar in the SDSS based on the presence of this component. 
The \civ\ absorption consists of at least five narrow components (see Fig~\ref{fig:abs_marked_fig}). Interestingly, the velocity separation between them is roughly consistent with the \civ\ doublet separation,
which suggests a possible
line-locking between these components.
We detect \nv\ absorption from all the individual components identified in the \civ\ trough.
We also detect \siiv\ absorption in some of our spectra (i.e epochs 1, 6, 7 and 9) from this system, in particular when \civ\ and \nv\ absorption equivalent widths are large. During these epochs, we also see additional \civ\ absorption covering the velocity range 8900 - 9900 kms$^{-1}$.
However, as this absorption is weak, we are unable to confirm whether these additional \civ\ absorption components are showing the same line-locking trend as seen for the above mentioned 5 components.

In our SALT spectrum, that extends to the bluer wavelengths (i.e till $\sim3200$\AA), we do see consistent broad absorption feature at the expected locations of \lya\ absorption.
However, it is blended with the \nv\ absorption from an intervening system at \zabs = 1.9098 (see Fig.~\ref{fig:abs_marked_fig}).
As seen in Fig~\ref{fig:abs_marked_fig}, the \nv\ absorption occurs on top of the broad \lya\  emission line. 
The \nv\ doublet absorption is completely blended and also contaminated by \lya\ absorption from the \zabs = 2.0457 narrow absorption system (see Fig~\ref{fig:abs_marked_fig}). 
It is clear that the \nv\ absorption has to cover some of the broad emission line photons as the residual flux is much deeper than the inferred power-law flux from our continuum fits.

\vskip 0.05 in
\noindent{\bf Covering factor:} We estimate the covering factor assuming the absorption lines to be saturated (a reasonable approximation for epoch-7, when the rest equivalent widths are maximum). In this case, the observed residual flux ($I_{obs}$) can be written as, $I_{obs}$ =  $(1-f_c)I_c + (1-f_e)I_e$,  where $f_c$ and $f_e$ are the covering factors with respect to continuum source and BLR region respectively and $I_c$ and $I_e$ are the unabsorbed continuum and broad emission line (BEL) fluxes as obtained from the continuum fits shown in Fig~\ref{fig:norm_spectra}.
 Both \civ\ and \nv\ absorption lines of BAL-A present flat cores between 7000-8300 kms$^{-1}$ in epoch-7 indicating saturated absorption.
We solve the above equation for \civ\ and \nv\ BALs assuming the absorption originate from the same region and obtain $f_c$ = 0.94 and $f_e$ = 0.75. 
Such an exercise is not possible for other velocity ranges as we are not sure about the line saturation. 

Thus, BAL-A component seems to not only cover the continuum emission region almost completely but also covers substantial area of the BLR. This suggests that the absorbing gas is either co-spatial or outside the broad emission line region.
 We estimate the size of the \civ\ BLR to  be $\sim$0.09 pc using the measured rest frame 1350 \AA\ monochromatic luminosity and equation 2 of \citet{kaspi2007}.
If we assume that individual absorption components have similar covering factor, then the size of individual clouds will be of the order of 0.07 pc.
High resolution spectra are needed to measure the covering factor and its variation  across the absorption profile using residual fluxes of the doublets for individual components at different 
epochs. 

\begin{figure}
    \centering
    \includegraphics[viewport= 40 30 1250 610,width=0.8\textwidth,clip=true]{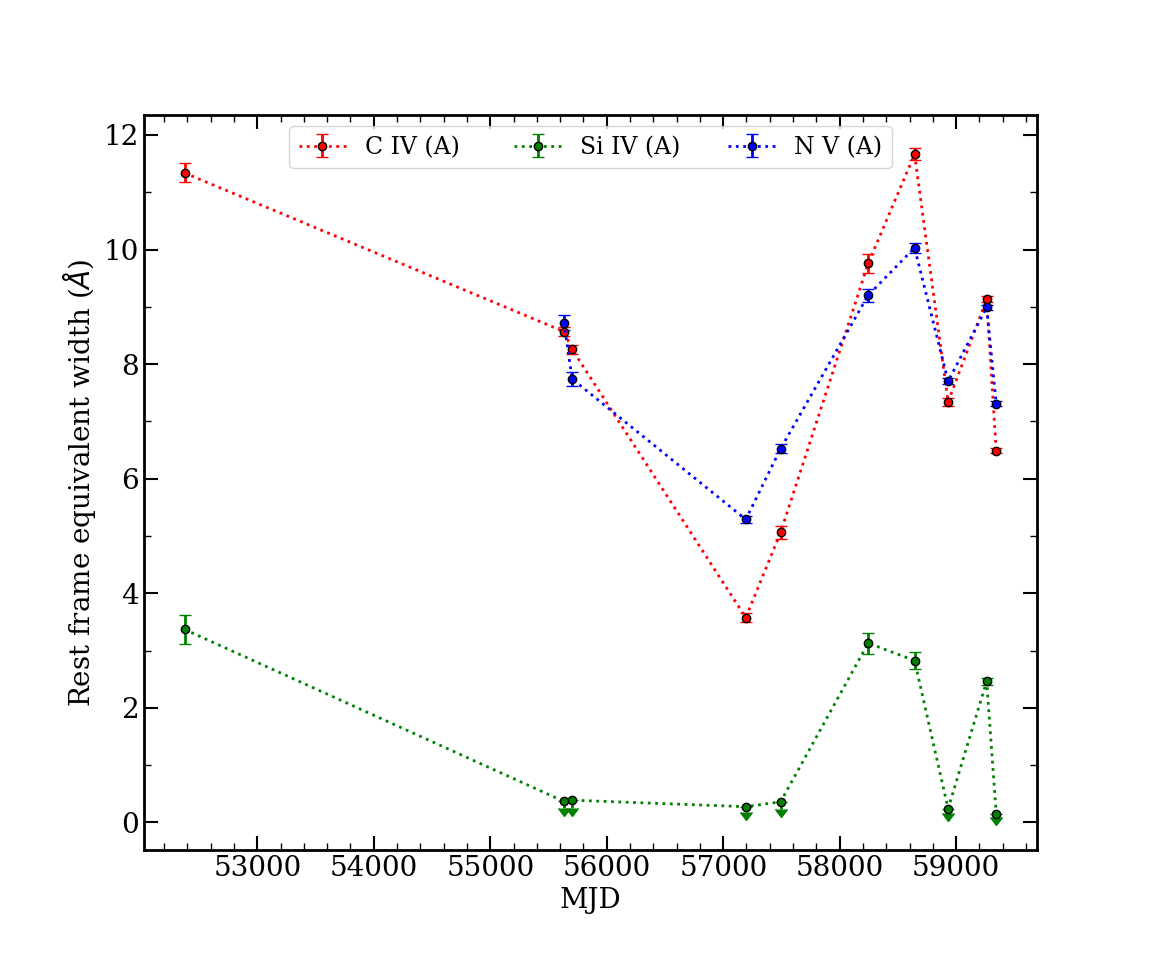}
    \includegraphics[viewport= 40 30 1250 610,width=0.8\textwidth,clip=true]{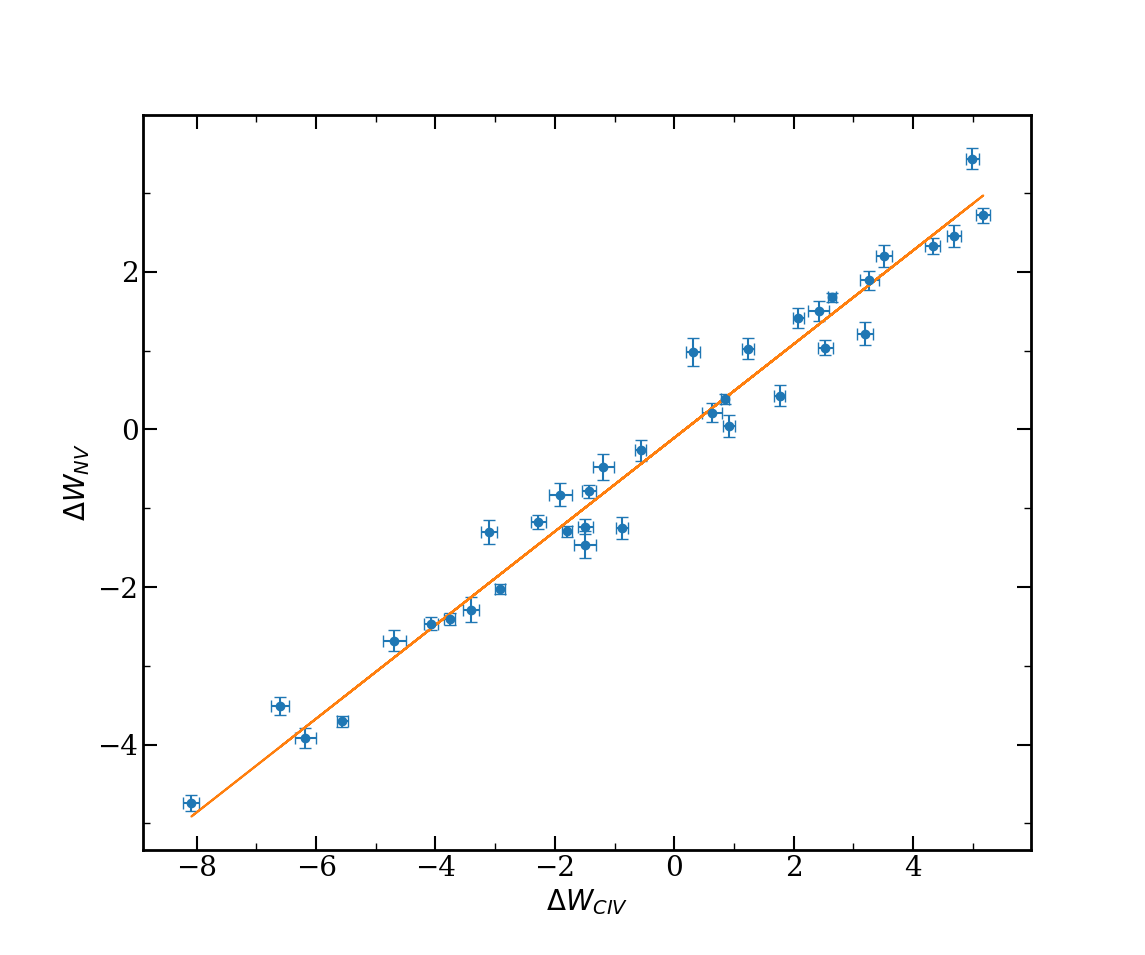}
    \caption{{\it Top panel}: Time evolution of \civ, \siiv\ and \nv\ rest equivalent widths for the BAL-A component. It is clear that the equivalent width of all the considered ions vary in unison. {\it Bottom panel:} Changes in the equivalent width of \civ\ are compared with that of \nv\ between all possible combinations of two epochs in our data. It is evident that changes in these two equivalent widths are well correlated and can be approximated by a linear line $\Delta W_{NV} = 0.58\times \Delta W_{CIV}$ - 0.01.}
    \label{fig:eqw_civ_A}
\end{figure}

\begin{figure}
    \centering
    \includegraphics[viewport=20 20 900 650, width= 0.5 \textwidth,clip=true]{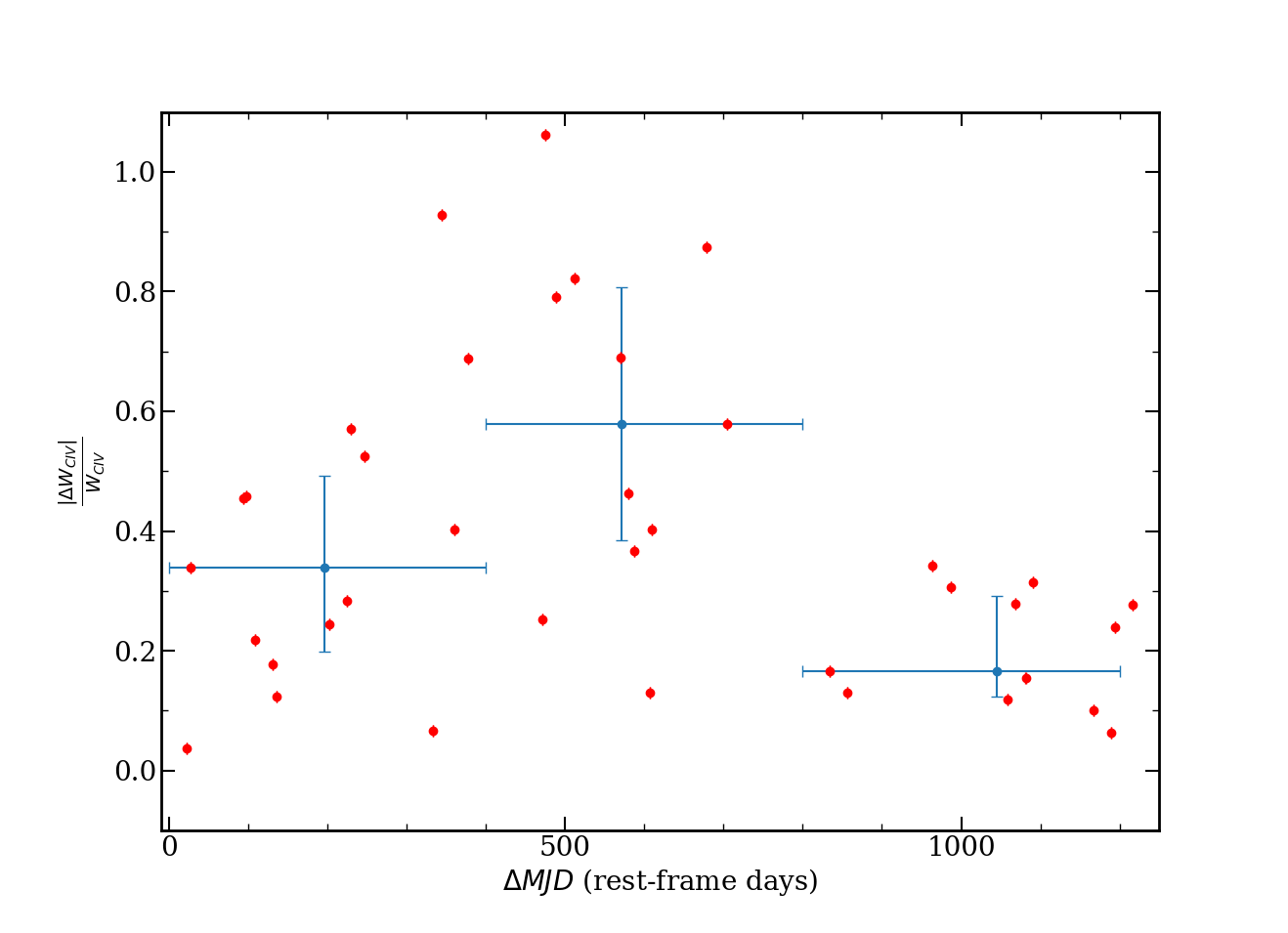}
    \caption{The fractional \civ\ equivalent width variations
    of BAL A component $\Big( \frac{| \Delta W_{C IV} | }{W_{C IV}} \Big)$ are plotted against the rest frame elapsed time ($\Delta$MJD). We used all possible pairs of observations.
    The median values of the same against the rest-frame time lags ($\Delta$ MJD) in bins of 400 days are shown as blue points with error bars. The upper and lower error bars correspond to the 75 and 25 percentiles.}
    \label{fig:delw_w_vs_time}
\end{figure}

\begin{figure}
    \centering
    \includegraphics[viewport=50 63 1035 1120,width= 0.49\textwidth,clip=true]{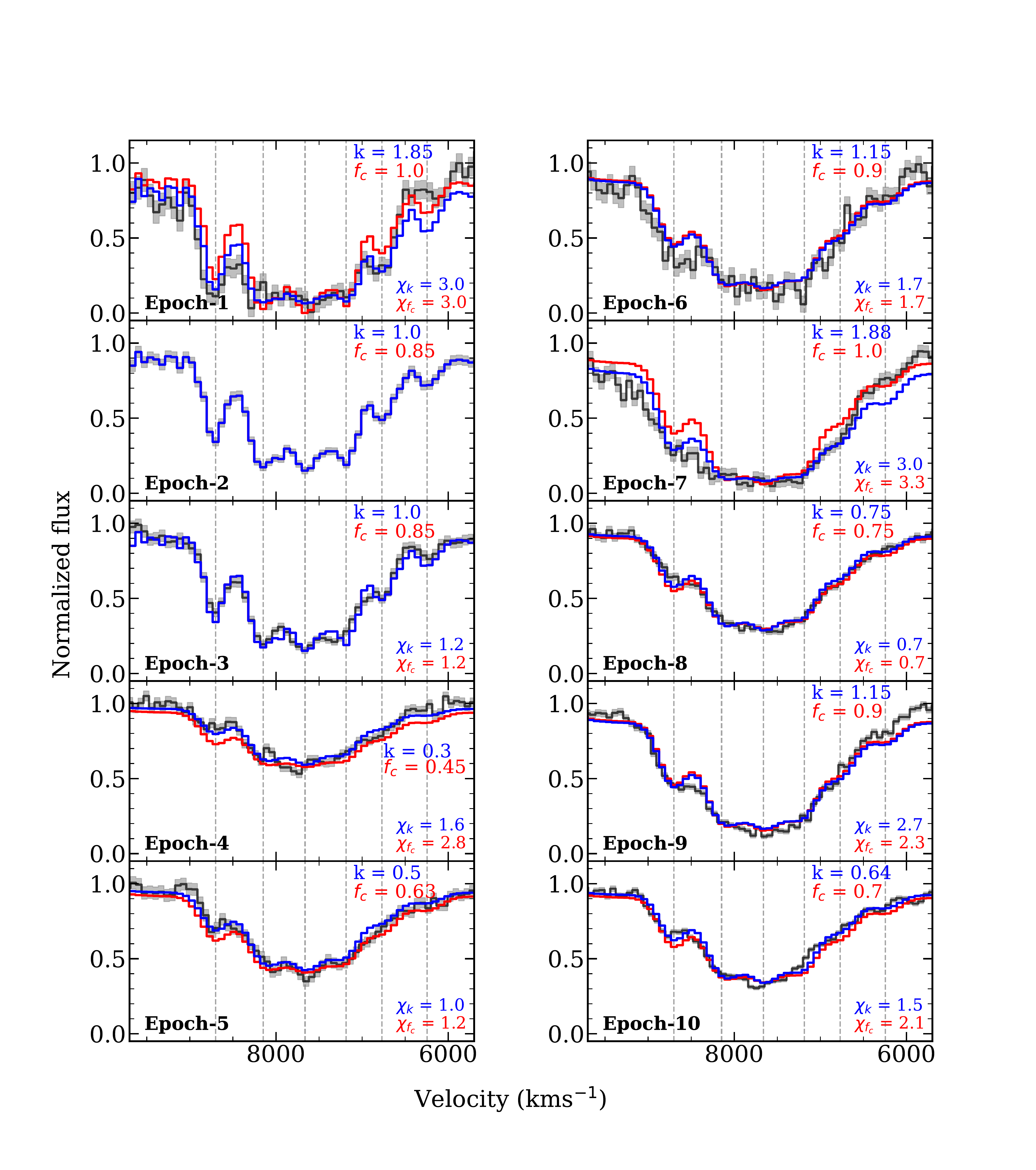}
    \caption{
    The \civ\ absorption profiles (plotted in units of ejection velocity with respect to \zem) of BAL-A observed during different epochs (black with flux errors shown in gray shaded regions). The red and blue profiles overlayed are predicted from the observed profile during epoch-2 after allowing for variations in covering factor and optical depth respectively (see text for details). The covering factor ($f_c$) or the scaling factor (k) applied to the optical depth 
    are indicated in each panel. Vertical dashed lines indicate the locations of individual \civ\ components seen in the SDSS spectra. Reduced $\chi^2$ for both cases ($\chi_{f_c}$ and $\chi_k$ for changing $f_c$ and optical depth respectively) are also given in each panel.
 }
    \label{fig:profile}
\end{figure}

\vskip 0.05in
\noindent{\bf The \civ\ variability time-scale}: From Fig.~\ref{fig:norm_spectra}, it is clear that the absorption lines of BAL-A show time variability. To quantify this, we measure the rest equivalent width of \civ, \lya, \siiv\ and \nv\ absorption of BAL-A component at each epoch using the normalized spectra. Measured \civ\ equivalent widths are summarized in Table~\ref{tab_obs} and  \lya, \nv\ and \siiv\ equivalent widths are summarized in Table~\ref{tab_bal_a}. Note that due to severe blending in  lower resolution spectra used here,  we could only estimate the total equivalent width that includes contributions from both the doublet transitions.
In the top panel of Fig.~\ref{fig:eqw_civ_A} we plot the measured \civ\ equivalent width as a function of MJD. From this figure and 
Table~\ref{tab_bal_a}, it is evident that \civ\ equivalent width has changed between any two epochs of observations.
To explore this further, we plot the absolute fractional change in  \civ\ equivalent width vs. the rest frame elapsed time ($\Delta$MJD) for all possible pairs of measurements in Fig.~\ref{fig:delw_w_vs_time}. 
The binned results (over 400 rest frame days shown using points with errobars) confirm the variation in the \civ\ equivalent width over different time-scales.
 The shortest time-scale (between epoch 9 and 10) over which significant rest equivalent width change (2.64$\pm$0.07\AA) seen is $\sim$27 rest frame days.
Largest variations are noticed over a time-scale of 400-700 rest frame days. 
 This could be due to variations either in the UV ionizing radiation or in the gas covering factor (probably due to transverse motions).
If we assume the first possibility then the recombination time-scale should be shorter than 
the smallest time-scale over which variations are seen. Using this arguement we obtain a 
lower limit on the electron density of 6 $\times 10^4$ cm$^{-3}$ for the assumed gas temperature of $10^4$ K \citep[using equation 9 of][]{Srianand2001}. 

\vskip 0.1in
\noindent{\bf Relationship between different species:} It is also apparent from Fig.~\ref{fig:eqw_civ_A} (top panel) that the variability in the rest equivalent width of all 
three ions considered here occurs in unison. 
To further explore this correlation we compare, in the bottom panel of Fig.~\ref{fig:eqw_civ_A}, 
the variations in the rest equivalent widths of \civ\ and \nv\ between all possible combinations 
of our spectroscopic epochs. These variations are well approximated by a straight line:  
$\Delta W_{\rm NV} = 0.58\times \Delta W_{\rm CIV}$ - 0.01.  This confirms that the \civ\ 
equivalent width varies more (i.e 1.72 times) than that of \nv. This could indicate 
that the \nv\ absorption is more saturated than the \civ\ absorption. 
The \siiv\ absorption is clearly detected when \civ\ and \nv\ rest equivalent widths are in 
excess of 9.1 \AA\ and 9.0 \AA\ respectively. 
From Table~\ref{tab_bal_a} we see that the rest equivalent width of \nv\ is less than 
that of \civ\ when \siiv\ absorption is detected. In
addition the \siiv\ equivalent width changes by more than an order of magnitude between 
epochs 1-2, 7-8 and 9-10. The corresponding reduction in \civ\ and \nv\ equivalent widths 
in the corresponding epochs is much smaller. This once again suggests the presence of 
line-saturation with \nv\ being more saturated compared to \civ.

\vskip 0.05in
\noindent{\bf Covering factor or optical depth variations? :}
It is natural to associate the equivalent width variations to changes in the optical depth 
(e.g. the column densities). However, 
discussions presented above suggest that \civ\ absorption is most likely to be saturated 
(at least when it is close to maximum). In such cases, changes in the covering factor also 
introduce large equivalent width changes even when there is no large variations in the ionising
radiation. Here, we first explore whether the observed \civ\ variations can be understood in the
framework of changes in the covering factor. We obtain the apparent optical depth as a function of ejection velocity (for epoch-2) using,
\begin{equation*}
    \tau(v) = -ln\bigg{(}{R(v)-1+f_c \over f_c}\bigg{)}.
\end{equation*}
Here, $R(v)$ is the measured residual flux at a given velocity. In the case of epoch-2, the centriods of \civ\ absorption from components 2-4 have similar residuals which under the assumption of saturation translate to $f_c\ge0.85$.
We construct the $\tau(v)$ for this epoch using a constant $f_c=0.85$, then we compute the \civ\ profile at other epochs by simply changing $f_c$ keeping $\tau(v)$ constant. Results of such an exercise are shown in Fig.~\ref{fig:profile} with red curves. In the case of SALT epochs, the profiles are convolved with an appropriate Gaussian to match the spectral resolution.
For epoch-1 spectrum, the absorption in the core are well reproduced by increasing $f_c$ to 1. Note that the predicted profile is not well reproduced in the wings. This is also the case for epoch-7 (when \civ\ is strongest). The core of the \civ\ absorption in epoch-4 spectrum is reproduced for $f_c = 0.45$. In this case the absorption in the wings are over-produced.
However, in the remaining cases, a simple scaling of the covering factor does produce reasonably good match (as suggested by the reduced $\chi^2$; i.e., $\chi_{f_c}$ in Fig~\ref{fig:profile}).
This exercise suggests that if there are no optical depth variations, then at least for the strongest absorption component, the covering factor is required to vary between 0.4 and 1.0.

Next, we consider the case where we keep $f_c$ constant over all epochs and allow  the optical 
depth to change by a constant factor through the profile.
We obtain the profile using,
\begin{equation}
    R_v = f_c~exp(-k \tau_v) + 1-f_c.
\end{equation}
Here $k$ is a scaling parameter that is varied to match profiles at different epochs. In this case we obtain $\tau_v$ from the second epoch spectrum assuming $f_c=0.95$ (close to what is found when the absorption is most saturated). The results for this case are shown with blue curves in Fig.~\ref{fig:profile}. This provide 
somewhat better fit (based on reduced $\chi^2$) to the observed profile at most epochs compared to the case with simple $f_c$ variations. However, we do see the fit is not perfect in the red wings in the highly saturated case. 
Between epoch-4 and epoch-7 (respectively epoch-1) the optical depth is required to vary by
a factor of 6 (respectively 7).

\vskip 0.1in
\noindent{\bf In summary}, BAL-A shows line-locking signature with \civ\ doublet splitting 
and strong absorption line variability over all measured time-scales. All the three ions 
detected show correlated variabilities. Large variability seen in \siiv\ (i.e close to an 
order of magnitude) equivalent width compared to that of \civ\ and \nv\ are consistent 
with the absorptions from the latter two species being saturated. Residual fluxes are seen 
in the core of \civ\ and \nv\ absorption even when their equivalent widths are maximum (i.e epoch-7). This suggests partial coverage for the absorbing gas which may be either 
co-spatial or outside the broad emission line regions. The observed variations 
of \civ\ absorption require a factor of $\sim$2 ($\sim$6) variation in the covering factor 
(optical depth) over the monitoring period. The optical depth variations provide slightly 
better fit to the data compare to pure covering factor variations. 
The overall velocity structure (whether there is any covering factor or 
optical depth change) remained nearly the same during our observing period. Such a stable
velocity structure in a line-locked  flow (to be confirmed at high spectral resolution) 
can place important constraints on the location of the flow \citep{srianand2000a}.

\begin{figure}
    \centering
    \includegraphics[viewport=20 20 900 650, width= 0.5 \textwidth,clip=true]{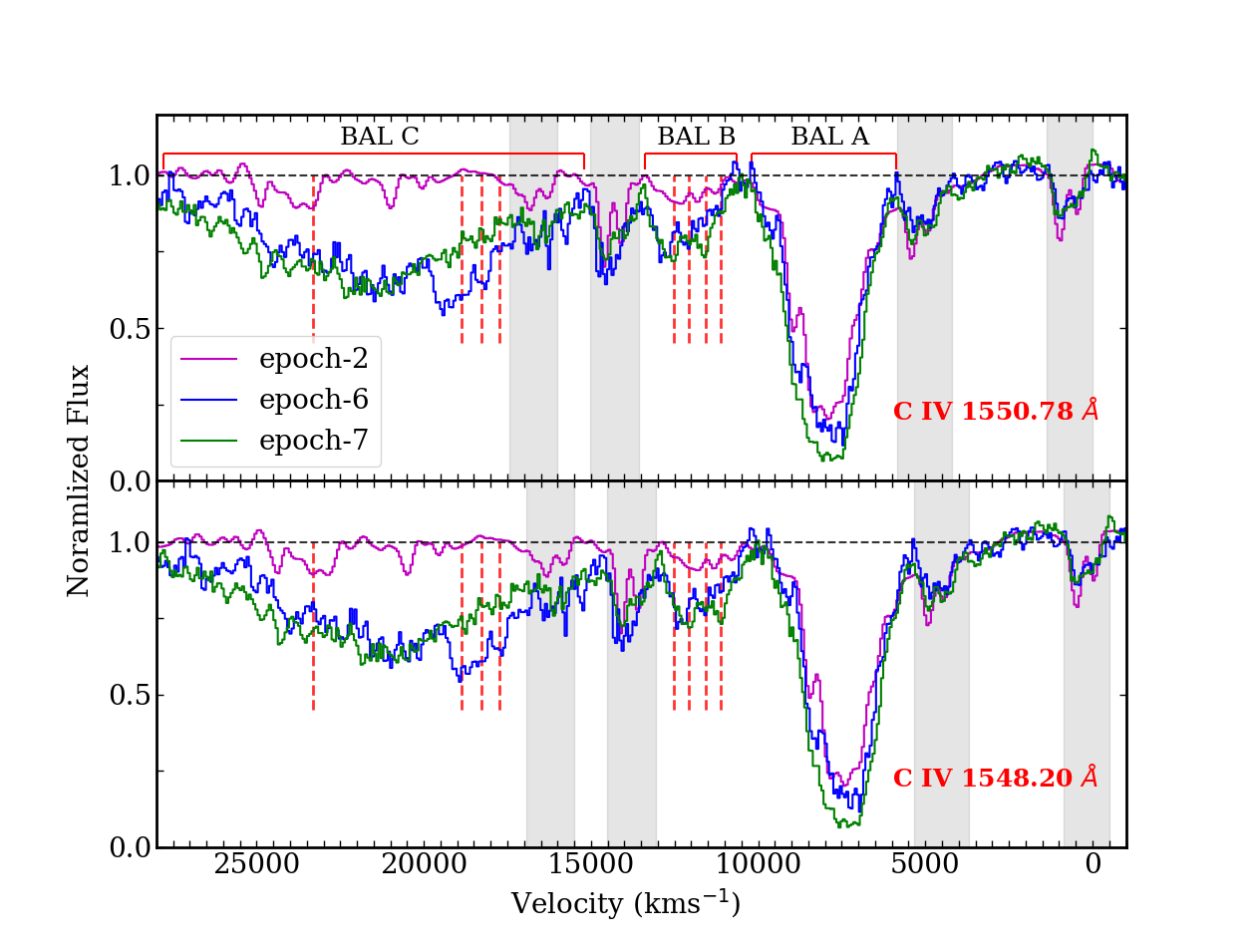}
    \caption{Normalised \civ\ profiles (during 3 epochs) in the velocity scale (with respect to \zem) covering the three BAL components of interest here. Shaded regions mark the narrow \civ\ absorption and vertical dashed lines identify the narrow \civ\ absorption that are part of the BAL components. 
    }
    \label{fig:clumps}
\end{figure}

\subsubsection{Observed properties of BAL-B :}

The \civ\ absorption from this component (spreading over the ejection velocities of 10500-13300 \kms) is clearly visible in the first epoch SDSS spectrum. \J13\ was added to our sample of quasars with ultra fast outflows for SALT monitoring due to the presence of this component . 
The measured rest equivalent width of 
\civ\ absorption at different epochs for this component is summarized in Table~\ref{tab_obs}. 
Fig.~\ref{fig:norm_spectra} and Table~\ref{tab_obs} suggest that the BAL-B variability is very similar to that of BAL-A. The maximum \civ\ equivalent width was measured 
during epoch-7.  As \nv\ and \lya\ absorption from this component are expected  in the \lya\ forest (see horizontal magenta lines in Fig.~\ref{fig:abs_marked_fig}), we could not probe the time variations of their equivalent 
widths.
We do not detect any \siiv\ absorption from BAL-B in any of our spectra.

The observed \civ\ profile during epoch-7 (when the equivalent width is maximum) can be well
approximated by 4 narrow components. These are consistent with two \civ\ doublets being present. The redshift difference between these two absorption are close to the \nv\ doublet splitting. This suggests the presence of line-locking as in the case of BAL-A.
Even in the first epoch spectrum we detect two distinct 
\civ\  doublets in the BAL-B profile that also have a velocity separation close to \nv\ 
doublet splitting. 
Interestingly, the location of these two components are different from 
what we find during epoch-7. 
 The variations in the depth of these components are visible in Fig.~\ref{fig:clumps} (vertical dashed line for BAL-B).
As the absorption lines are relatively weak we need high resolution spectra to confirm 
the existence of line-locking and possible kinematic shift between the strongest components. 

\subsubsection{Observed properties of BAL-C :}
\label{sec:BALC}
This BAL component was not present in the  spectra observed during the first three epochs.
However, it is evident from Fig.~\ref{fig:norm_spectra} that this component (BAL-C; spread over 15000-29000 kms$^{-1}$) has emerged, grew to reach a maximum equivalent width during epoch-6 and subsequently weakened during our SALT monitoring period.
As the \civ\ absorption is shallow and spread over a large velocity range, continuum fitting is important to confirm its presence and to measure its rest equivalent width accurately. 
\begin{figure}
    \centering
    \includegraphics[viewport=55 50 1050 1250,width= 0.49\textwidth,clip=true]{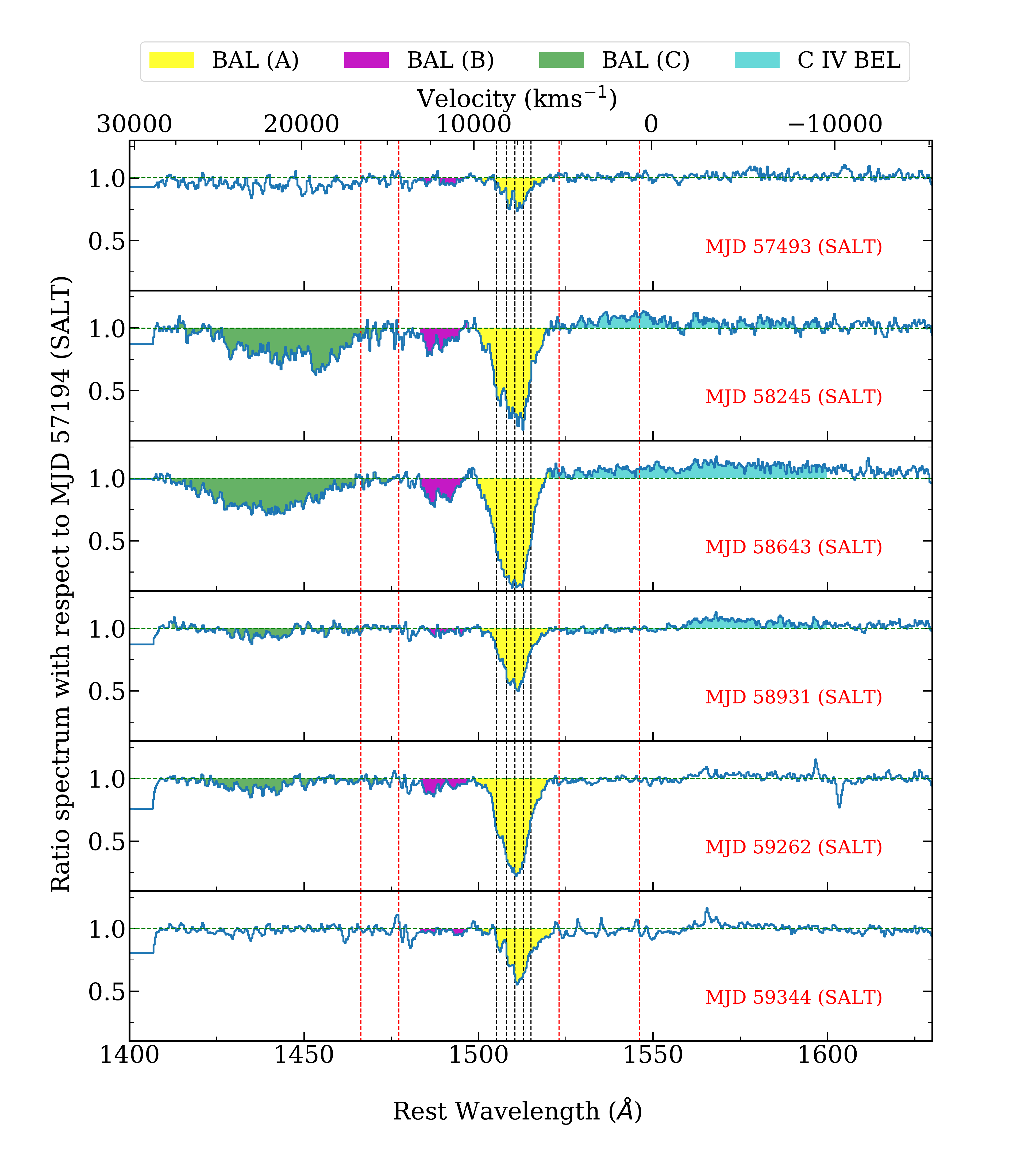}
    \caption{Ratio of the SALT spectra obtained during the last 6 epochs with respect to the epoch-4 SALT spectrum. Any deviation from the dotted horizontal line downwards  corresponds to absorption line variability. The red vertical dashed lines identify the locations of the narrow \civ\ absorption systems. The black dashed lines indicate the location of the five distinct components in BAL-A. This figure confirms the increase followed by steady decrease in the apparent \civ\ optical depth of component-C. The region shown in cyan also indicates possible variability in the \civ\ broad emission line. }
    \label{fig:ratio_spec_all}
\end{figure}
As can be seen from Fig.~\ref{fig:abs_marked_fig}, the \nv\ absorption from this component is blended with \lya\ absorption from  narrow \civ\ absorption systems at \zabs = 1.9098 and 1.8888. It is possible that this part of the spectrum is also contaminated by intervening \lya\ forest absorption. Thus it is difficult to accurately measure the \nv\  equivalent width at different epochs.

To capture the variability of \civ\ absorption, avoiding continuum fitting issues, in Fig.~\ref{fig:ratio_spec_all} we show the ratio of SALT spectra (around the range of different 
\civ\ BAL components) obtained in last 6 epochs with respect to the epoch-4 spectrum. As in Fig.~\ref{fig:abs_marked_fig}, different absorption components in the ratio spectrum (that correspond to absorption line variability) are shown in different colours.
This figure confirms the trend we notice in Fig.~\ref{fig:norm_spectra}. The \civ\ rest equivalent width measured from the normalised spectra are also summarized in Table.~\ref{tab_obs}. 

\begin{figure}
    \centering
    \includegraphics[viewport=0 0 850 650, width=0.47\textwidth,clip=true]{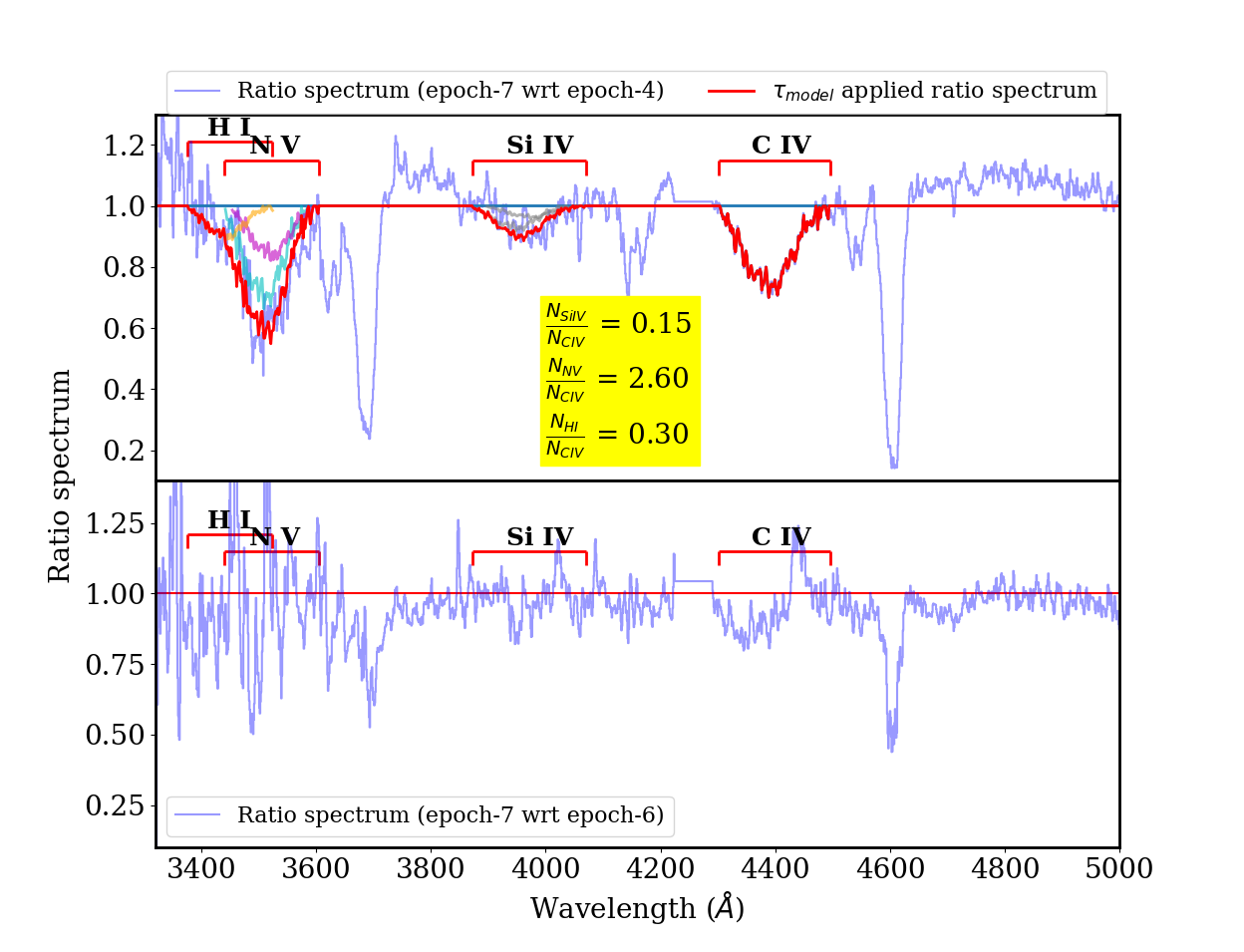}
    \caption{{\it Top:} the observed ratio spectrum of epoch-7 with respect to epoch-4 .The predicted absorption for various ionic species (red) given the observed \civ\ BAL C profile and the ion-fraction vs U relation from CLOUDY photo-ionization models are shown (see Section~\ref{sec:discuss} for details).{\it Bottom:} Ratio spectrum of epoch-7 with respect to that of epoch-6 where the BAL component shows minor variations.}
    \label{fig:abs_model}
\end{figure}

\begin{figure}
    \centering
    \includegraphics[viewport=50 20 1050 1190, width=0.5\textwidth,clip=true]{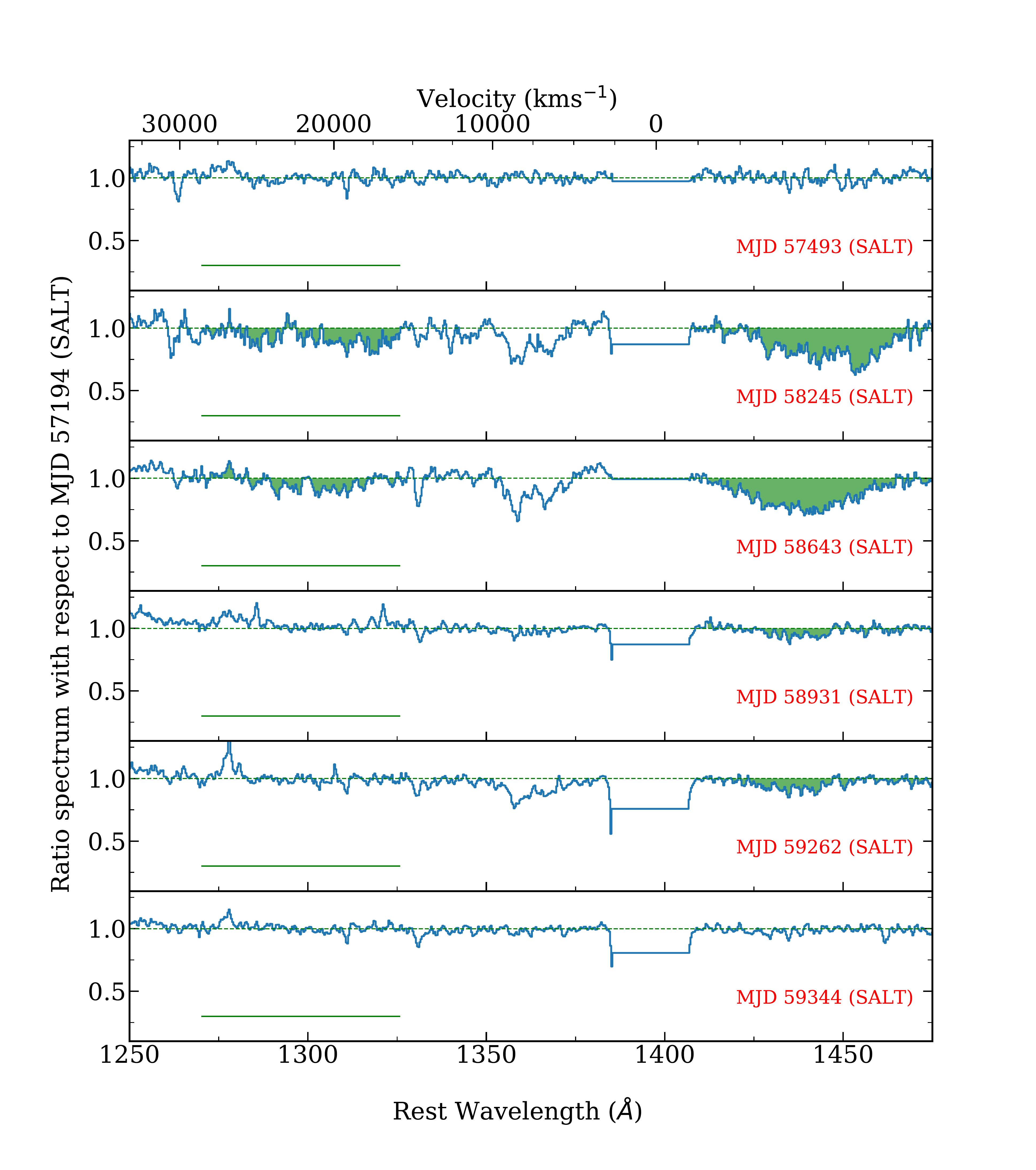}
    \caption{Same as Fig.~\ref{fig:ratio_spec_all} but covering the Si~{\sc iv} range (green horizontal lines) of BAL-C. Detectable Si~{\sc iv} optical depth variations are seen only during epoch-6 and 7.  The velocity scale for \siiv\ absorption with respect to the systemic redshift (\zem = 2.0498) is provided at the top.
    }
    \label{fig:si4ratio}
\end{figure}

Next to explore the equivalent width variations of \nv\ and \siiv\ absorption,  we plot the ratio of the spectrum obtained for epoch-7 with respect to the epoch-4 (and epoch-6) spectrum in Fig.~\ref{fig:abs_model}. The strong variation in the \nv\ absorption is clearly visible. Even in this case the changes in the rest equivalent width of \civ\ and \nv\ happen in the same direction as we found in the case of BAL-A. We also see possible dip in the ratio spectrum at the expected position of 
\siiv\ absorption. This suggests that, like in the case of \civ\ and \nv, the equivalent width of \siiv\ also shows increasing trend. To confirm the \siiv\ variability we show the ratio plot in the \siiv\ region in Fig.~\ref{fig:si4ratio}. Possible dip is seen in the expected position of \siiv\ absorption during epoch 6 and 7 (second and third panel from the top). We do not see any non-zero variation for other epochs.

Even though the overall absorption profile of BAL-C looks smooth, a closer look at the profile reveals interesting features. During epoch-1 while the broad absorption is not clearly visible we notice narrow absorption features around
the ejection velocity of 21,360 \kms.
Similarly in the epoch 2 spectrum we notice extra absorption consistent with being a \civ\
doublet at an ejection velocity of 23,800 \kms (see Fig.~\ref{fig:clumps}). We also see some narrow absorption feature around ejection velocity of 18,500 \kms in the spectrum obtained in epoch-6 (see Fig.~\ref{fig:clumps}).
These absorption features (that are present only during one epoch) can not be associated with any of the narrow \civ\ systems discussed before.
Interestingly all these features are consistent with being \civ\ absorption. Thus it is possible that the emerging component BAL-C 
encompasses rapidly varying narrow features.
This could indicate the presence of rapidly evolving clumps of gas embedded in an overall smooth flow. 

The large ejection velocities coupled with large velocity spread are consistent with the emerged gas being 
close to the central engine (i.e region of high gravitational potential) and associated with the accretion 
disk-wind  \citep{arav1994a, murray1995}. As can be seen from Fig.~\ref{fig:abs_marked_fig}, the absorption mainly occurs against the 
continuum emission i.e.,  against the accretion disk emission and not BEL. Therefore, it is difficult to identify the location of the absorbing gas relative 
to the BLR through covering factor arguments as we have done in the case of BAL-A.
We put  a lower limit on the covering factor ($f_c$) of BAL-C after considering the minimum transmitted flux 
along the \civ\ BAL profile. This turns out to be around $<$70 $\%$ of the total continuum flux for \civ\ BAL-C 
implying $f_c \geq 0.30$ (as the lines need not be saturated). This means that the projected size of this
flow is very small.

\begin{figure}
    \centering
    \includegraphics[viewport=50 30 800 620, width=0.47\textwidth,clip=true]{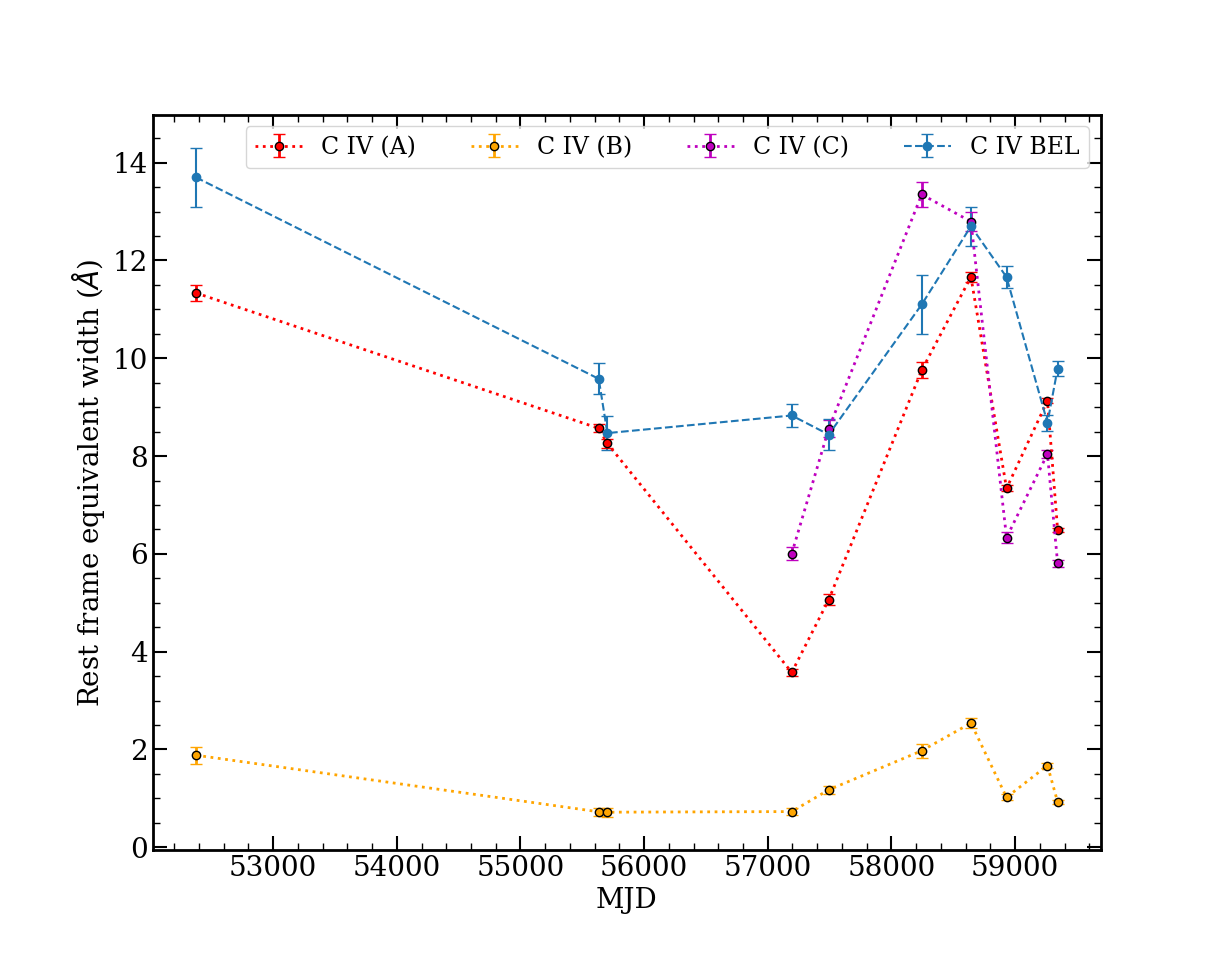}
    \caption{Time  evolution in the rest equivalent width of the \civ\ absorption of BAL-A, B and C components. 
    The evolution of \civ\ emission line equivalent width is also shown. The correlated variability between 
    different BAL components and BEL are clearly seen.}
    \label{fig:eqw_civ_all}
\end{figure}

\subsubsection{Correlated BAL variability}

In Fig.~\ref{fig:eqw_civ_all}, we compare the time evolution of C~{\sc iv} equivalent widths of different identified BAL components. For both BAL-A and BAL-B, \civ\ rest equivalent widths were minimum during epoch-4, then showed a sharp increase and reached a maximum during epoch-7 before declining.
The strong correlation between BAL-A and -B equivalent widths is confirmed by the Spearman-rank correlation analysis (with r = 0.7 with a p-value of 0.01). From Table~\ref{tab_obs}, it is clear that the fractional change in equivalent width between any two epochs is larger for BAL-B compared to BAL-A.  
This could just be related to BAL-A component being more saturated than BAL-B if the entire variability is driven by variations in the photo-ionization. However, recall the possible component variations in the case of BAL-B discussed above. This needs to be considered while interpreting the correlated variability. 

It is also evident from Fig.~\ref{fig:eqw_civ_all} that the time evolution of  BAL-C, while following overall trend, is slightly different. For example, the peak \civ\ equivalent width occurs during epoch-6 for this component whereas for other components, the peak is during epoch-7.
It is also evident from Table~\ref{tab_obs}, that  \civ\ equivalent width of BAL-C during epochs 4-7 is higher than that of BAL-A. But it becomes weaker than that of BAL-A in the last three epochs.
The Spearmann rank correlation analysis suggests a strong correlation (i.e r = 0.79 and p-value of 0.04) between BAL-A and BAL-C considering the last 7 SALT epochs data.
The slight deviation shown by BAL-C can be attributed to the narrow components that appear and disappear between the two epochs (see the discussions in Section~\ref{sec:BALC}).

 Correlated \civ\ absorption line variability between components at different velocities has been reported in the literature using large samples of quasars \citep[e.g.,][]{capellupo2011,Filiz2013,wang2015, rogerson2018}. 
For example, \citet{Filiz2013} have found similar fractional changes in the \civ\ equivalent width between low and high velocity BAL troughs. They  argued that the correlated \civ\ absorption line variability is due to changes in the shielding gas properties (i.e changes in the \civ\ ionizing radiation).
The correlated variability we find in the case of \J13\ (using 10 epoch spectra covering a wide range of time-scales) is consistent with their findings.

\begin{figure}
    \centering
    \includegraphics[viewport=105 20 1200 605, width=0.47\textwidth,clip=true]{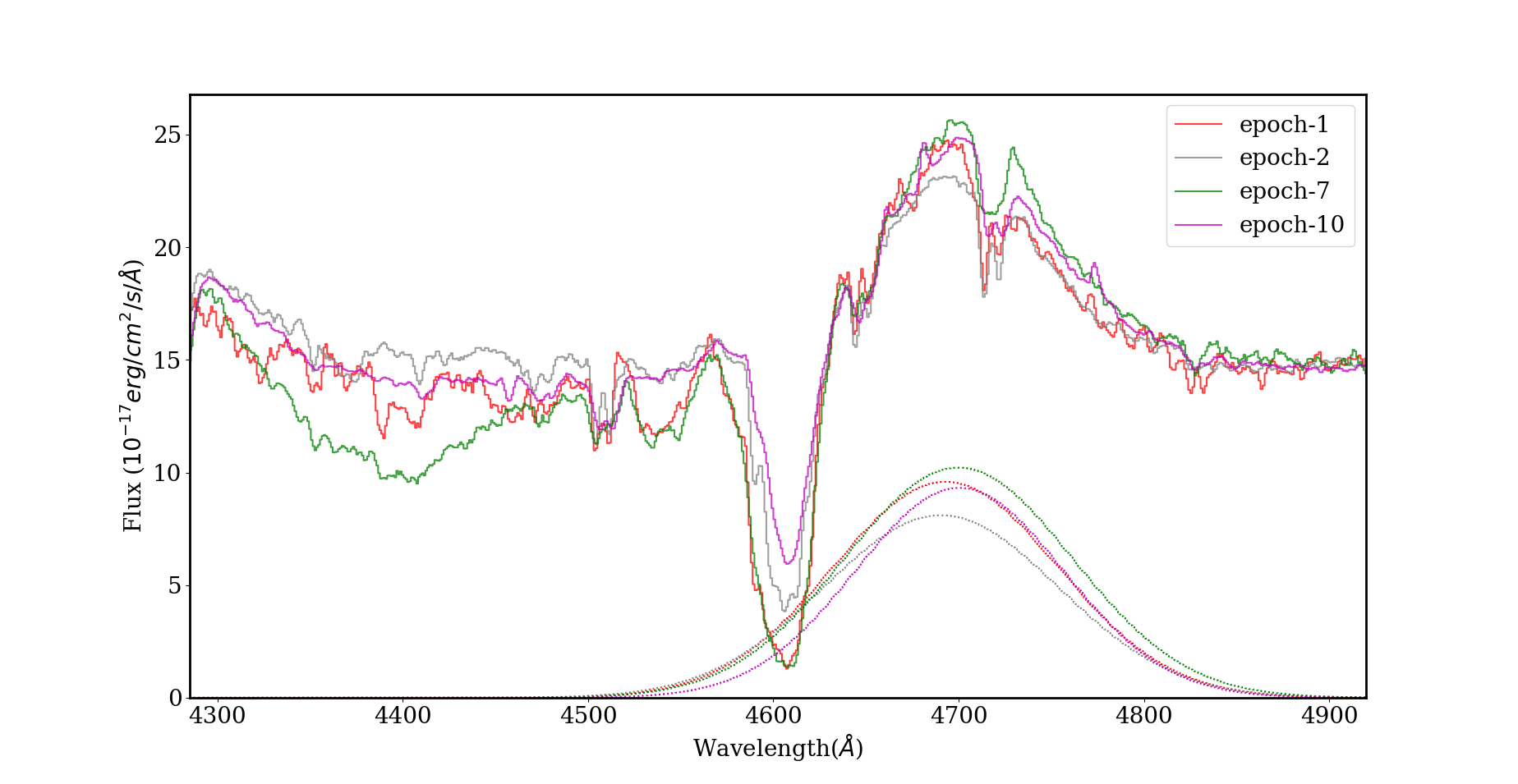}
    \caption{Observed \civ\ emission line profile together with the best fitted Gaussians at each epoch are also provided.
    The \civ\ emission line is weaker during epoch-2 (and 3) compared to other epochs. }
    \label{fig:bel}
\end{figure}

We show the time evolution of the \civ\ emission line rest equivalent width obtained from our continuum+Gaussian fits in Fig.~\ref{fig:eqw_civ_all}. Interestingly this follows the overall 
trend set by different BAL components with subtle differences.
Compared to epoch-1, epoch-2 spectrum shows less \civ\ absorption equivalent widths and
we notice the \civ\ emission line flux is less during this epoch as well 
(see Fig.~\ref{fig:bel}). The \civ\ absorption equivalent widths grows and reaches 
maximum value during epoch-7. The \civ\ emission line flux follows the same trend. 
In the subsequent epochs, the \civ\ absorption and emission equivalent widths decrease. 
The observed variations of the \civ\ emission profile over short time-scales are consistent 
with being driven by ionization induced variations (i.e reverberation). In that case 
it suggests possible variations in the \civ\ ionizing flux over our monitoring period. 

It is usual procedure to interpret such a correlated variability between different BAL
components and BEL equivalent widths to changes in the ionization due to quasar variability
\citep[see for example,][]{aromal2021}.  \citet{wang2015} have found a correlation between variations in the equivalent widths of \civ\ emission lines and BALs in their sample. They also concluded that the absorption line variations are driven by ionization changes.  
Covering factor variations will not be able to explain the observed \civ\ emission line variations (that samples the full volume) unless otherwise one introduces large changes 
in the gas distribution in the broad line region (that can only take place over larger
time-scales, i.e typically over the sound-crossing time). 

\section{Photometric variability of \J13}
\label{sec:continuum}
\begin{figure}
    \centering
    \includegraphics[viewport=10 20 600 620,width=0.45 \textwidth,clip=true]{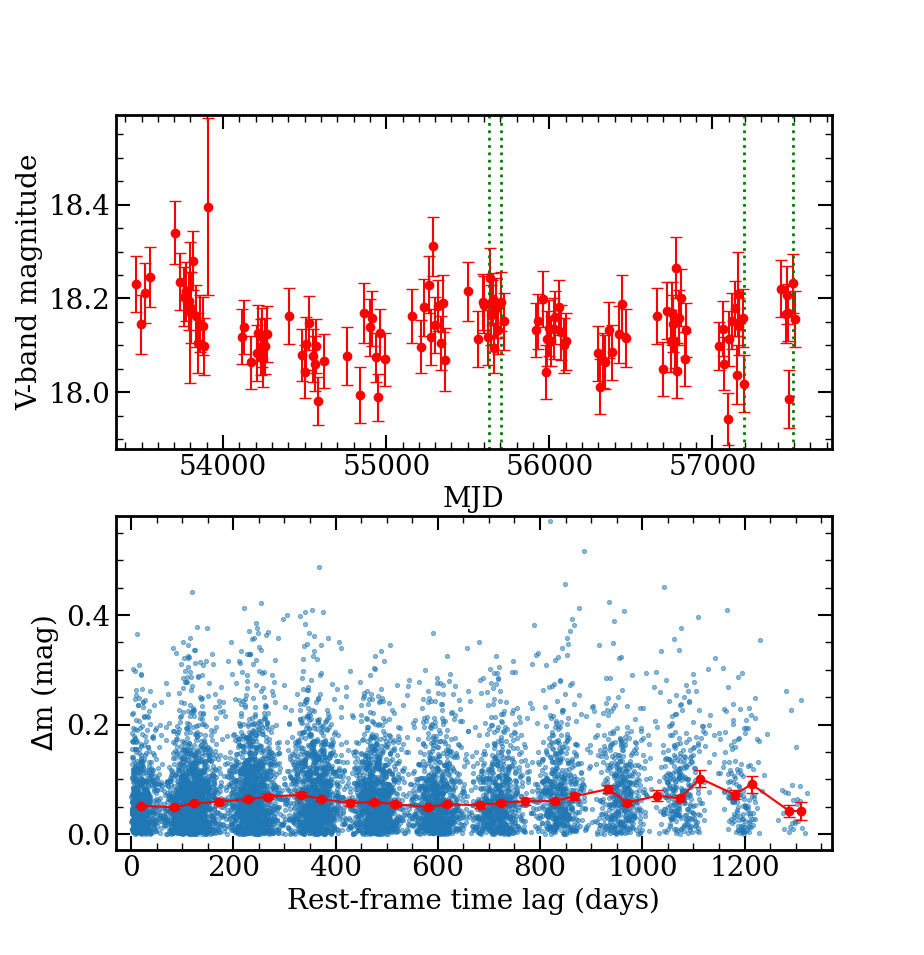}
    \caption{{\it Top panel:} V-band light curve of \J13 from CRTS. The vertical dotted lines indicate the four spectroscopic epochs (epochs 2 to 5). {\rm Bottom panel:} Structure function obtained from the light curve shown in the top panel. The red points are the average values in a rest frame time-lag bin of 50 days.}
    \label{fig:crtslc}
\end{figure}

\begin{figure}
    \centering
    \includegraphics[viewport=15 30 760 650,width=0.47 \textwidth,clip=true]{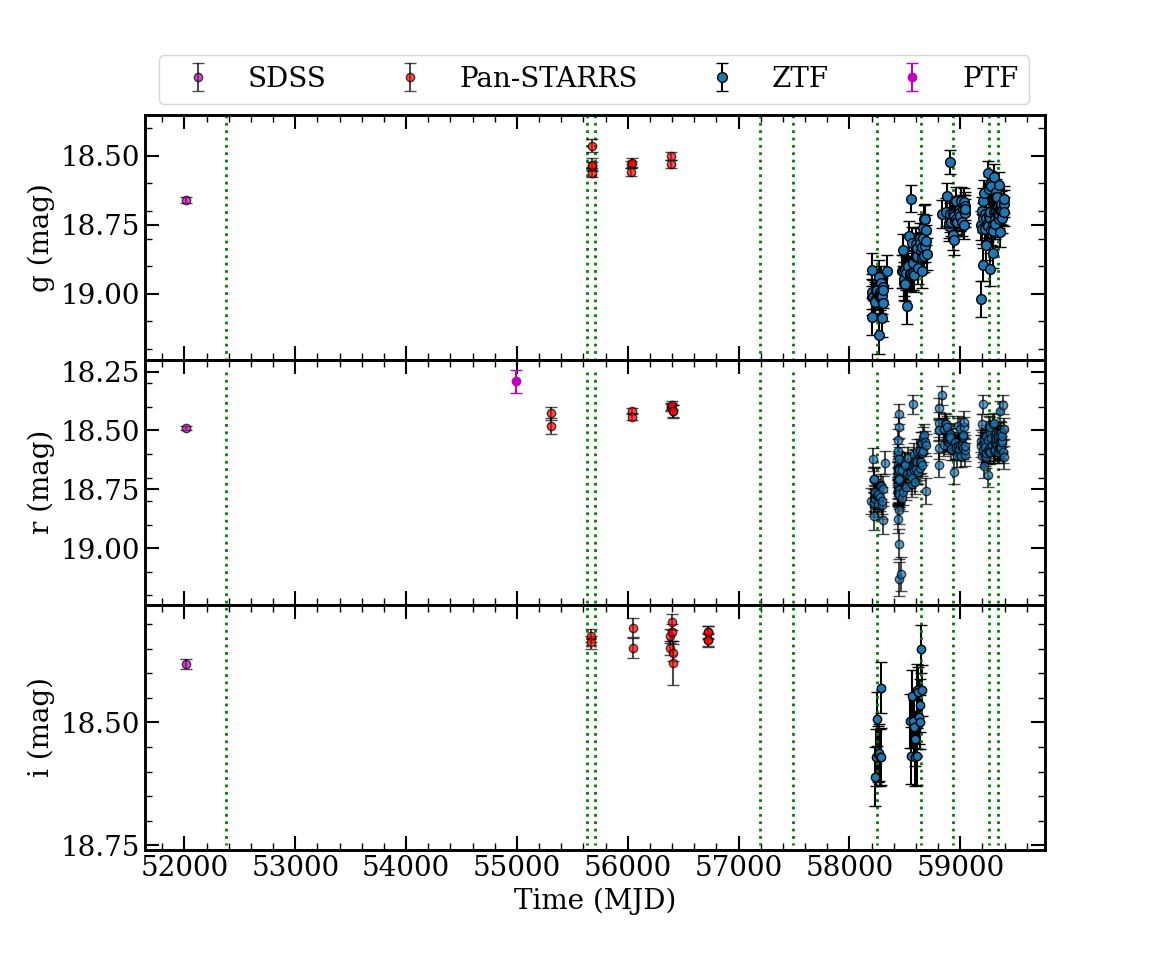}
    \caption{g, r and i-band light curves from SDSS, Pan-STARRS and ZTF. Different spectroscopic epochs are identified by the vertical dashed lines. These light curves together with the CRTS light curve (shown in Fig.~\ref{fig:crtslc}) suggest a possible dimming of \J13\ between epochs 5 and 6 following by brightening by up to 0.3 mag in subsequent epochs.}
    \label{fig:full_light_curve}
\end{figure}

In Fig.~\ref{fig:crtslc}, we plot the 
CRTS V-band light curve of \J13\ in the top and the measured structure function \citep[as
defined in][]{MacLeod2012} in the bottom panels. This covers our spectroscopic epochs 2 to 5
(shown as the green dashed lines). There is no systematic photometric variation (fading or brightening beyond 0.1 mag over large time-scales) 
on any time-scale. During this period the \civ\ rest equivalent width of BAL-A has reduced 
by a factor of 2.4 between epochs 3 and 4. Clearly the rest equivalent width variations 
are much larger than the fractional variation in the quasar magnitude seen in the CRTS 
light curve.
It is well known that CRTS light curves (obtained without any filter), while sensitive to
overall variations in the quasar brightness, will not capture colour variations accurately.

In Fig.~\ref{fig:full_light_curve}, we plot the publicly available light curves of \J13\
obtained in three different filters (g, r and i-bands).  While these light curves are not
sampled uniformly in time,  ZTF measurements provides good coverage during our last 5
spectroscopic epochs.
It can be seen from this figure that \J13\ was faint around epoch-6 compared to SDSS and
Pan-STARRS photometric points and ZTF measurements in subsequent epochs. This together with 
the nearly flat CRTS light curve over the first five epochs seen in Fig.~\ref{fig:crtslc} are
consistent with a fading of \J13\ between epochs 5 and 6. It is also clear that the quasar has
brightened between epoch 6 and 7 and remained nearly at constant magnitude level for the last 3
epochs (within 0.05 mag in g- and r-bands).
However, the absorption equivalent width variations are not monotonic as one would expect 
from the light curves.
It is clear from Table~\ref{tab_obs} that \civ\ rest equivalent widths of all the three
components have shown significant correlated variability (by 25 to 40\% for BAL-A; by 45 to 64\%
for BAL-B and 27 to 28\% for BAL-C) over the period covered by the last three epochs where 
the g and r-band light curves are flat.

We have ZTF photometric measurements available within two days of our SALT spectroscopic
measurements for the last 5 epochs. We used them to fix the absolute flux scale of our spectra
as described in \citet{aromal2021}. Including three SDSS spectra, we have 8 epochs for which
spectra with good flux calibration are available. We use these to further probe the correlation
between \civ\ equivalent with and continuum flux variations.

\begin{figure}
    \centering
    \includegraphics[viewport=15 25 890 630,width=0.48 \textwidth,clip=true]{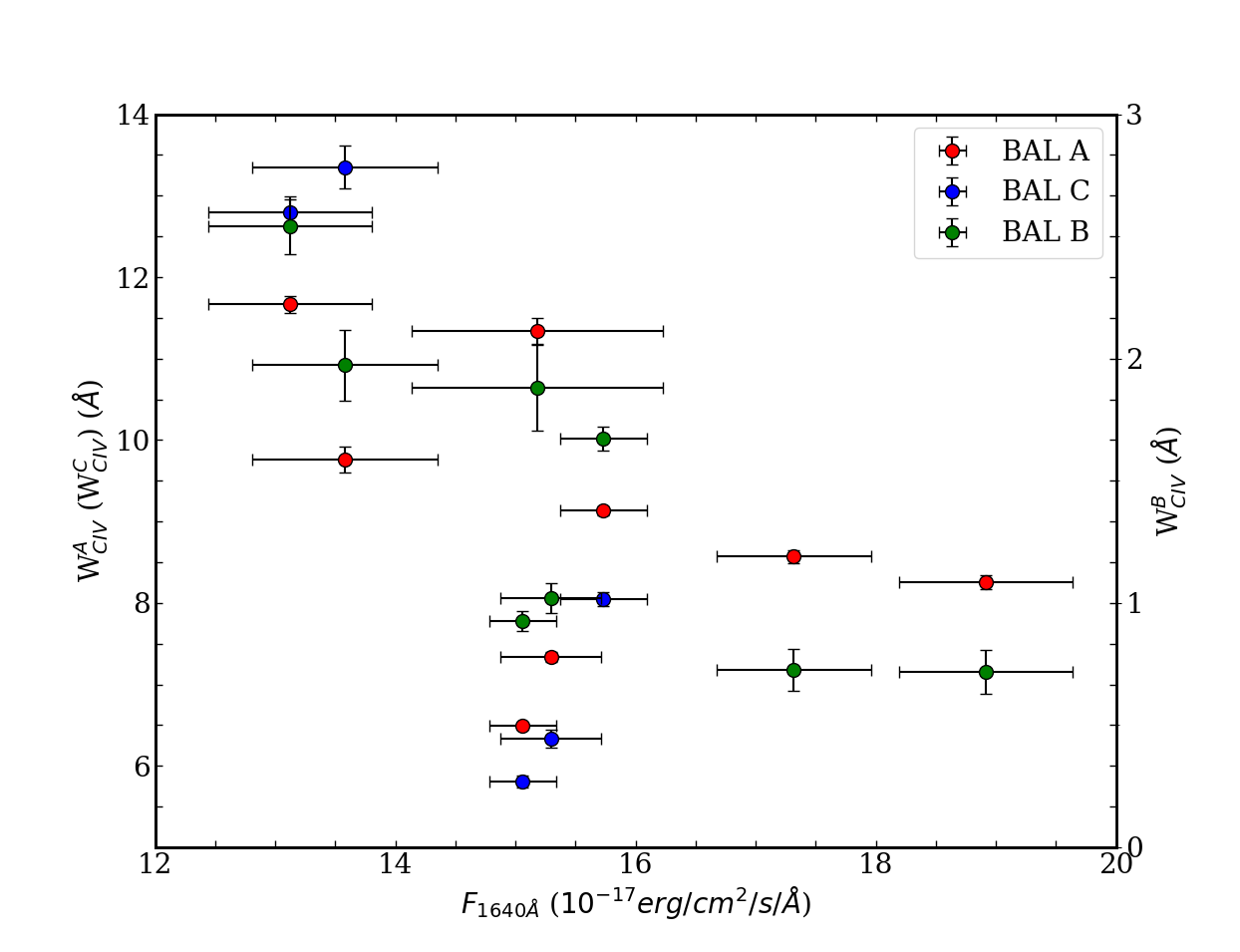} 
    \includegraphics[viewport=0 0 860 630,width=0.47 \textwidth,clip=true]{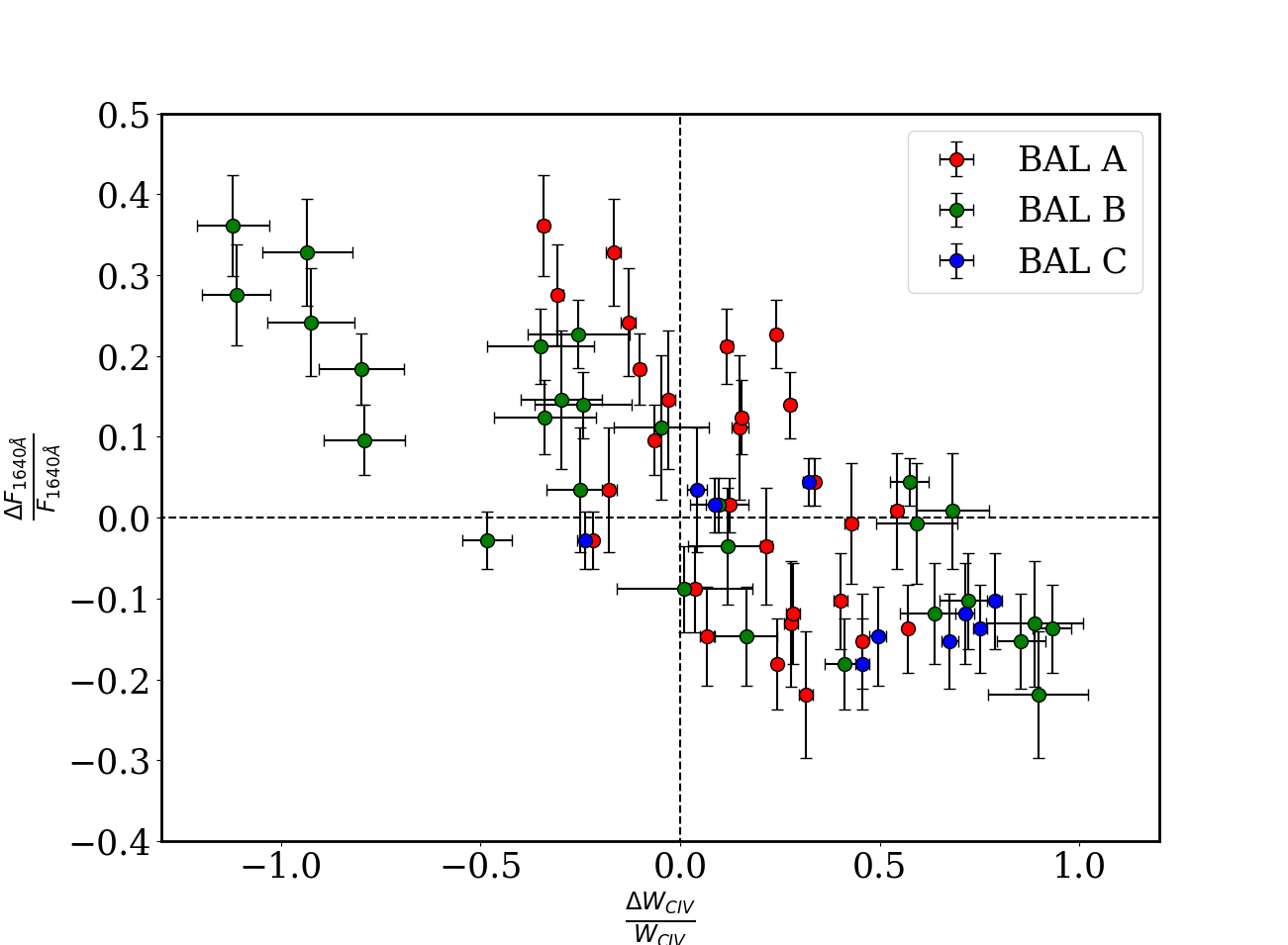}
    \caption{
    {\it Top:} Continuum flux at rest frame 1640\AA\ vs \civ\ rest equivalent width of 
    different BAL components studied here. A correlation is clearly present between these
    quantities for all components. Such a correlation strongly favors photoionization
    induced variability. 
    {\it Bottom:} Fractional change in the \civ\ equivalent width vs. continuum flux measured. Results are presented for different BAL components. A correlation is clearly
    visible (albeit with large scatter) between the two quantities. The percentage variation in
    the continuum flux is found to be smaller than what has been seen for the equivalent widths.
    In addition we see variations in the equivalent widths between epochs with no strong
    flux variations.
    }
    \label{fig:del_eqw_vs_del_flux}
\end{figure}

We quantify the continuum flux using the monochromatic flux at observed 5000 \AA (i.e at the
rest frame wavelength of 1640\AA), $F_{1640 \text{\AA}}$,  measured from our flux calibrated
spectra. Between the first two SDSS epochs $F_{1640 \text{\AA}}$ has increased by a factor 1.15
while the \civ\ equivalent width has reduced by a factor 1.32.  $F_{1640 \text{\AA}}$ was nearly
the same between epochs 6 and 7. However, we do see the \civ\ equivalent width being increased
by a factor 1.20.

In the top panel of Fig.~\ref{fig:del_eqw_vs_del_flux}, we plot the \civ\ rest equivalent width
against $F_{1640 \text{\AA} }$. All the BAL components show a clear anti-correlation between the
\civ\ rest equivalent width and $F_{1640 \AA}$. Such a trend is usually interpreted as an effect
of changes in the photoionization equilibrium.
However, for a given $F_{1640 \text{\AA}}$ there is a large spread in
\civ\ equivalent width. In the bottom panel of Fig.~\ref{fig:del_eqw_vs_del_flux}, we plot the
fractional variation in \civ\ equivalent width against the fractional variation in $F_{1640
\AA}$ between all possible pairs of spectra. There is a strong correlation between the two
quantities. In the case of BAL-A component (Spearman rank correlation r = -0.61 with a p-value of $5\times10^{-4}$), we notice a spread in $\Delta EW/EW$ of $\sim0.5-0.7$ at any given $\Delta
F/F \sim 0$. As noted before, BAL-B shows a stronger anti-correlation (Spearman rank
correlation r = -0.85 with a p-value of $7\times10^{-9}$) and  much larger fractional variation in \civ\ equivalent width compared to BAL-A. 
This could be related to BAL-A being more saturated
compared to BAL-B (as apparent from Fig.~\ref{fig:abs_marked_fig}).
In the case of BAL-C, as it was not present during the three SDSS epochs we do not have
measurements in the 1st quadrant. Hence, the observed correlation is relatively weak (r = 
-0.5 and p-value of 0.15). Looking at the 1st and 4th quadrant (in the bottom panel of
Fig.~\ref{fig:del_eqw_vs_del_flux}) it is evident that a similar fractional change in rest
equivalent width is associated with larger changes in the flux in the 1st quadrant (i.e
increasing flux and decreasing EW) compared to the 4th quadrant (i.e decreasing flux and
increasing EW).
Such a trend is expected when (i) the absorption lines are near saturation that can lead to
asymmetric change in equivalent with for similar change in column density (induced by the
changes in the ionizing flux) around the mean,
and/or (ii) the mean ionization parameter is close to when the \civ\ fraction is maximum which can also lead to asymmetric response to change in ionizing flux.  

The requirement of large changes in the ionizing flux can be reconciled if we allow for the
\civ\ ionizing photon flux to have larger variability compared to what we measure in the FUV
range (i.e $F_{1640 \text{\AA}}$).  The scatter can be understood if the \civ\ ionizing photon
flux has a range of values for a given $F_{1640 \text{\AA}}$ \citep[similar discussions in][in
the case of J1621+0758]{aromal2021}. On the other hand, it is also possible that in addition to
the changes in the ionizing flux, variations in other parameters such as the covering factor
(see discussions in the previous section) or changes in the gas properties could  
contribute to the scatter. The changes in the line of sight gas distribution is also 
suggested in the case of BAL-B and BAL-C through the presence of rapidly varying narrow
absorption components as discussed above.

\citet{Misra2021} have found the newly appeared BALs are associated with dimming of the 
quasars. Interestingly, the emergence of BAL-C happens close to the dimming episode in the
lightcurve of \J13 (see Fig.~\ref{fig:full_light_curve}).
\citet{Misra2021} interpreted the appearence of a new BAL component as the consequence of 
lower ionization. Alternatively, it is possible that both the dimming and BAL emergence can be
related to changes in the accretion disk structure \citep[see][for discussions in the case of
J1333+0012]{Vivek2012}. In the following section we analyse the observed variability properties
in the framework of simple photoionization models.

\section{Discussions}
\label{sec:discuss}
\subsection{Photoionization model}

\begin{figure}
    \centering
    \includegraphics[viewport=20 150 600 700, width=0.5\textwidth,clip=true]{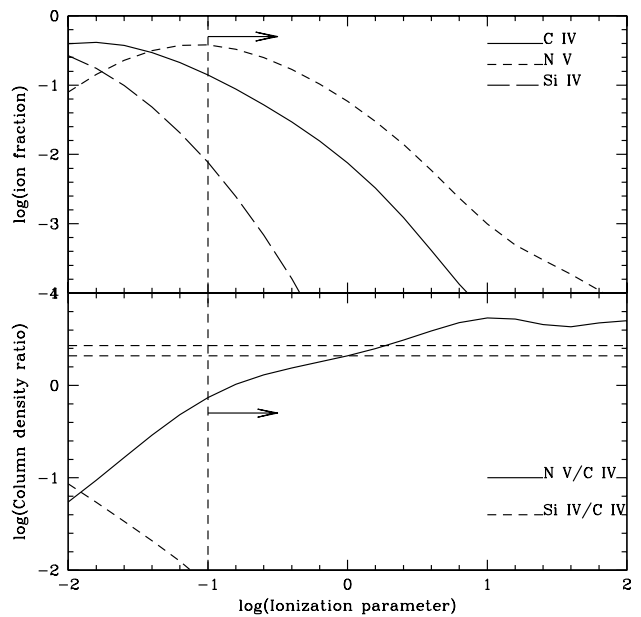}
    \caption{{\sl Top}: Ion fraction as a function of ionization parameter. The vertical dashed line in this panel marks the ionization parameter above which both \nv\ and \civ\ change in unison. {\sl Bottom}: Column density ratio as a function of ionization parameter.  }
    \label{fig:PI}
\end{figure}

In this section, we try to understand the \civ\ absorption line variability of BAL components in the framework of simple photo-ionization (PI) models. Based on the inferred covering factor (for epoch-7 when absorption lines are strongest) it is most likely that the absorbing gas is either co-spatial with or located outside the broad emission line region (typical distance of $\sim 0.09$pc from the central UV source).
First we construct the broad band spectral energy distribution (SED) of \J13 using the method discussed in \citet{aromal2021} (refer to subsection 5.2.1) using the available photometric data. Then, we use the code {\sc Cloudy} to simulate the thermal and ionization conditions in the absorbing gas (assumed to be a plane parallel slab of uniform density with solar metallicity) after illuminating it with the above SED from one side. We run {\sc Cloudy} v17.01 \citep{cloudy2017} under optically thin (to the \hi\ ionizing photons) conditions and vary the ionization parameter (U) from log~U = $-$5 to 2 in steps of 0.2 dex. For the assumed SED the number of ionizing photons per second from \J13 is, Q $\sim 2.7\times 10^{56}$ s$^{-1}$. In this case the distance of the absorbing gas ($r$), ionization parameter (U) and number density ${\rm n_H}$ are related  by,
\begin{equation}
{r \rm = 139.3 \times \bigg{(}{10^5\over n_H}\bigg{)}^{0.5}~~ \bigg{(}{10^{-1.4}\over U}\bigg{)}^{0.5} }~~~{\rm pc}.
\label{eqn:u}
\end{equation}
Constraints on the models come from the measured correlation between the \civ\ and \nv\ equivalent widths and the anti-correlation between the fractional change in equivalent width and continuum flux. 
 Ideally one needs to compare the model predicted column densities with the observed ones. Also the single slab considered here may not be an ideal representation of multiple components seen in observations. Thus, we do not make any rigorous attempt to model all observational aspects discussed in the previous section. But draw some general conclusions.

\subsubsection{PI model for BAL-A}

In the top panel of Fig.~\ref{fig:PI}, we show the ion fraction as a function of the ionization parameter. In order to reproduce the observations that both \civ\ and \nv\ absorption show correlated variability we need the ionization parameter to be greater than 0.1 (vertical dashed line at log~U=$-1$ in the top panel). In this regime the column density of both \civ\ and \nv\ will increase with decreasing quasar flux as suggested by our observations (see Fig.~\ref{fig:del_eqw_vs_del_flux}).
To detect \siiv\ absorption, the ionization parameter should be closer to log~U =$-1$.
In this ionization parameter range very small fraction (i.e $<1\%$) of Si remains as \siiv.
Our models suggest that the
\siiv\ column density will be $\sim$ two orders of magnitude smaller than that of \civ\ (see bottom panel of Fig.~\ref{fig:PI}). Thus PI models suggest a strong saturation for \civ\ and \nv\ during epochs when we detect \siiv\ absorption. Ideally the \nv\ and \civ\ equivalent widths should be same in the case of strong saturation as they only probe the velocity spread. The small difference we find in their equivalent width ratios indicates possible differences in the covering factors of \civ\ and \nv.

We have noted that between epoch 7 and 8 (or 9 and 10) the \siiv\ equivalent width has changed by more than a factor 10. This can be produced by an increase in the ionization parameter by a factor of 2 (i.e a decrease by 0.75 mag). Such a change will produce a factor of 2 reduction in the \civ\ column density while we observe a factor of 1.6 reduction in the \civ\ equivalent width. Our measured $F_{1640 \text{\AA}}$ has changed by 13\% between epoch 7 and 8. This confirms our earlier suggestion that for photo-ionization induced variations to be effective we require a larger change in the number of \civ\ ionizing photons compared to the number of UV photons at 1640\AA.

As discussed before the recombination time-scale less than 26 days gives a lower limit on $n_H$ of $\ge 6\times10^{4}$ cm$^{-3}$ \citep[using equation 9 from][]{Srianand2001}. This together with the lower limit of log~U=$-1.0$ suggests $r<113$ pc from the central ionizing source.

\subsubsection{PI model for BAL-C}
As BAL-C component also shows variations of \civ\ and \nv\ in the same direction the constraint on log~U will be similar to what we have for BAL-A (i.e log~U$>-1$). In the top panel of Fig~\ref{fig:abs_model}, we plot the ratio of spectra obtained during epochs 4 and 7. In the absence of absorption line variability we expect the ratio spectrum to be flat. At the locations of the absorption lines the ratio spectrum gives the change in the apparent optical depth. Using the \civ\ optical depth difference profile for BAL-C we are able to reproduce its the \nv\ profile by a simple scaling (i.e  d$N$(\civ) = 0.38$\times$ d$N$(\nv)).
This exercise, shows that the ratio of \civ\ to \nv\ column density is nearly constant across the profile.

To obtain more insights, we considered the \civ\ apparent optical depth profile measured during epoch-7 using the continuum normalised spectrum.
Using this we predict the \nv\ absorption profile for various log~{U} assuming same $U$ for the full velocity range (we investigate this assumption bit more in the next section).
The profiles at the expected position of \nv\ for log~U=0.2 (avoiding the regions contaminated by narrow \lya\ absorption) matches well with the predicted profile.
At this ionization parameter we expect very little \siiv\ absorption. Note the ionization parameter will be lower if we allow for [N/C] ratio to be more than solar as found in the case of associated absorbers \citep[][]{Petitjean1994}. We notice that even when we allow for a factor of 3 higher abundance of N with respect to C the models will not produce sufficient \siiv\ absorption. Thus the possible presence of \siiv\ absorption seen in Fig.~\ref{fig:abs_model} does not arise naturally in the PI model.
This could mean one  or more of: (i) \civ\ and \nv\ absorption being saturated and estimates based on apparent optical depths underestimate their column density, 
(ii) non-solar metal abundance of Si (i.e excess by an order of magnitude) with respect to N and O and/or (ii) multi-phased outflows well mixed in the velocity space so that there are no clear profile differences between different ions. 

The option (i) is the simplest explanation. However the \civ\ absorption line does not show flat bottom. This could just mean multi-streaming flow having different covering factor over different velocity ranges. 
As noted before the large velocity dispersion shown by the \civ\ absorption is inconsistent with single slab model.
In the following, we see whether these issues can be addressed using simple density and velocity profiles of the absorbing material.

\subsection{BAL-C and disk wind}
\label{sec:disk_model}

\begin{figure}
    \centering
    \includegraphics[viewport=40 15 800 630,width=0.48 \textwidth,clip=true]{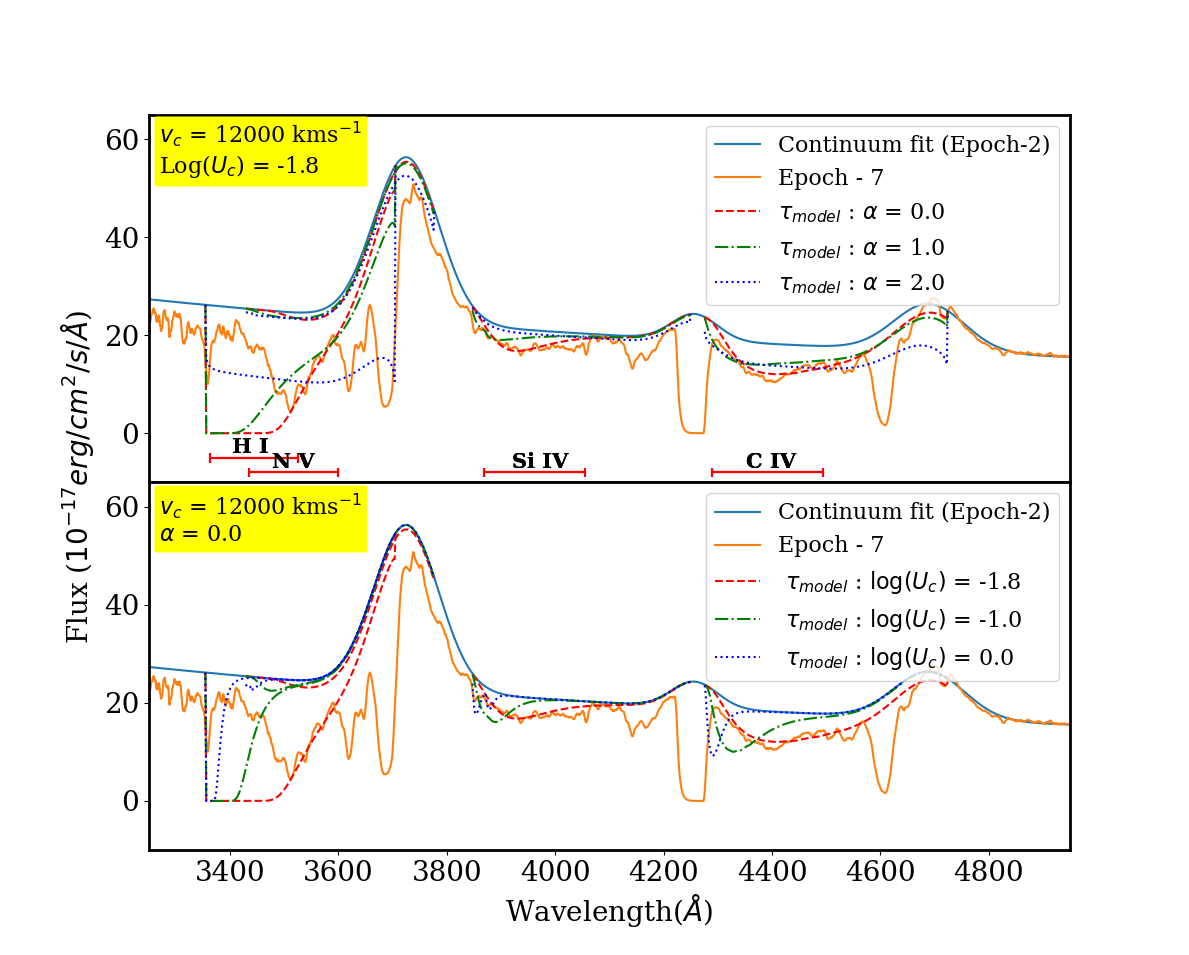}
    \caption{{\it Top panel:} Observed spectrum of \J13\ during epoch-7, when BAL-C has the maximum \civ\ equivalent width. The unabsorbed spectrum (the best fitted continuum, shown in blue, to epoch-2 spectrum when BAL-C was not present.)
    is also shown. The predicted absorption profiles of \civ, \siiv, \nv\ and \lya\ absorption (using the density and velocity law described in section~\ref{sec:disk_model} and {\sc Cloudy} models) for $v_c$ = 12000 \kms, ${\rm log~U_c}$ = -1.8 and three values of $\alpha$ are shown.
    {\it Bottom panel:} The effect of varying ${\rm log~ U_c}$ at a given $v_c$. The model with large ${\rm log~ U_c}$  tends to shift the peak \civ\ absorption towards higher velocities. Also note none of the models that reproduce the observed \civ\ profile produce sufficient \nv\ absorption and all of them over produce the \lya\ absorption.
    }
    \label{fig:wind_model}
\end{figure}


\begin{table}
\begin{threeparttable}
    \centering
\caption{Model parameters}
 \begin{tabular}{cccc}
  \hline
   Parameter & $\alpha$=0  & $\alpha$=1 & $\alpha$=2 \\ 
  \hline
  \hline
  $v_{\infty}$ (kms$^{-1}$) & 30000 & 30000 & 30000 \\
  $r_f$ (pc) & 0.02 & 0.02 & 0.02 \\
  $n_0$ (cm$^{-3}$) & 2 $\times 10^7$ & 5 $\times 10^5$ & 5 $\times 10^4$ \\
  $r_0$ (pc) & 1 & 1 & 1 \\
  $v_c$ (kms$^{-1}$) & 12000 & 12000 & 12000 \\

  \hline
 \end{tabular}

\label{tab_parameters}
\end{threeparttable}
\end{table}

As pointed out before, the large velocity width of BAL-C absorption could mean the absorbing gas is associated with a disk-wind. Simple 1D disk-wind models predict velocity and density as a function of radial distance from the accretion disk. Here, we investigate whether the absorption profiles of BAL-C are consistent with such a scenario.
We create a model absorption profile for the BAL-C component by incorporating {\sc Cloudy} simulation results with an assumed radial density and velocity profile. Wind models predict a radial velocity profile of the form \citep[see][]{murray1995},
\begin{equation}
    v(r) = v_{\infty} \Big(1-\frac{r_f}{r} \Big)^\beta 
\end{equation}
where $r$ is the distance to the cloud from the central source, $v_{\infty}$ is the terminal velocity, $\beta \sim 1.15$ and $r_f$ is the launching radius of the wind. 
In the literature it is usual to assume $r_f$ to be the distance at which the circular velocity is $v_\infty$. We also assume a density law, $n(r)$, as follows,
\begin{equation}
    n(r) = n_0 \Big( \frac{r}{r_0} \Big)^{-\alpha} 
\end{equation}
where $n_0$ $ and $ $r_0$ are constants and $\alpha = 2$ consistent with a mass-conserving spherically expanding wind. %
We consider density structures corresponding to three values of $\alpha$ (0 [constant density], 1 and 2).
The ionization parameter (U) of the gas is given by, 
\begin{equation}
    U=\frac{Q_H}{4 \pi n_H c r^2}
\end{equation}
where $Q_H$ is the total number of hydrogen ionizing photons produced by the source, $n_H$ is the density of hydrogen in the cloud and $c$ is the speed of light. 
The value of $Q_H$ is fixed by the assumed quasar SED and the measured continuum fluxes.
Once $v(r)$ and $n(r)$ are normalised with an appropriate choice of constants in equations 2 and 3, $U(r)$ and hence the ion densities at different velocities are given at any radial distance.
In all our models we consider $v_\infty = 30,000$ \kms\ and $r_f=0.02$ pc which fixes $v(r)$. We then assign $log~U_c$ at a given velocity $v_c=12,000$ \kms. For a given density law this fixes the ionization parameter of the gas at any velocity $v$. The value of $n_0$ is fixed to match the maximum observed optical depth of the \civ\ line. 
The 
column density of \civ\ originating from any  distance interval ($\Delta r$) at a distance $r$ can be obtained using,
\begin{equation}
    N_{theory}(r) = n_H(r) [C/H] f_{C IV}(r) \Delta r
\end{equation}
where [C/H] is carbon metallicity (assumed to be solar) and $f_{C IV}$ is the \civ\ ion fraction.  Similarly we can predict the column density profile and hence the absorption profile of ions of our interest.
 The \siiv\ absorption profile predicted by the model ($v_c$ = 12,000 \kms and ${\rm log~U_c}$ = -1.8) are also consistent with observations.
Since the density and velocity profiles are monotonous in these models, low \civ\ optical depth at high velocities is produced by gas with low ionization parameter (i.e \civ\ ion fraction is decreased by going to lower ionization parameter at large distances/velocities). This gas produces strong \lya\ absorption contrary to what has been observed. Also these models under-produce the \nv\ absorption. 
We were unable to get consistent fits to the observed \civ\ and \nv\ absorption just by changing the values of different parameters. In Fig.~\ref{fig:wind_model} we illustrate the effect of varying $\alpha$ and ${\rm U_c}$. In the top panel of Fig.~\ref{fig:wind_model} we show the effect of varying $\alpha$ for a fixed $v_c$  and ${\rm U_c}$. The model with $\alpha = 0$, while reproducing the \civ\ and \siiv\ profiles well over-produces the \lya\ absorption and under-produces the \nv\ absorption. Increasing the value of $\alpha$ makes the absorption profile flat and broad, inconsistent with the observed profiles.
In the bottom panel of Fig.~\ref{fig:wind_model}, we show the effect of increasing the ${\rm log~U_c}$ for a fixed $v_c=12,000$ \kms. It is evident that the peak optical depth in \civ\ absorption occurs at larger velocities for larger values of ${\rm log~U_c}$.
This is mainly because ${\rm log~U}\sim-2$ is now shifted to larger distances and velocities.
However, all these models produce less \nv\ absorption compared to what is observed.
We notice that changing $r_f$ does not improve the situation. In the case of $\alpha=1$, we find that the profile shape can be made to match the observed profile by changing $r_f$ values. However, as in the case of $\alpha=0$ models even these models do not produce sufficient \nv\ absorption because of the reasons discussed above.
Thus it appears that smooth density and velocity profiles do not produce either the observed profile or the ion ratios correctly.

In MHD simulations, the radial distribution of density and velocity depends on the inclination angle of our line of sight with respect to the disk plane \citep[see for example figure 2 of][]{Dyda2018a}. It will be interesting to check whether \civ, \nv, \lya\ and constraints on \siiv\ absorption of BAL-C can be reproduced from such simulations \citep[using similar exercise to][that focus on very highly ionized species]{Ganguly2021}.

\section{Summary}
\label{sec:summary}

Using the Southern African Large Telescope (SALT) we have been carrying out a systematic study of 
time variability of BAL absorptions in a sample of 63 quasars from SDSS DR15 \citep{paris2017} which show BALs at outflow velocities greater than 15,000 kms$^{-1}$ (UFOs) in their spectra.  
In this paper, we present a  detailed analysis, J132216.25+075808.4 ($z_{em}$ = 2.0498), that shows a strong correlated variability in absorptions spread over 6000-29000 \kms. Such a variability is rare
and also not noticed among other objects in our sample.
For ease of discussions, we identify three BAL components well separated in the velocity space (namely BAL-A, B and C in the ascending order of the outflow velocity).  

Our analysis suggests that \civ\ BAL-A, at outflow velocities of 5800 - 9900 \kms, consists of multiple narrow absorption features that are separated roughly by the \civ\ doublet splitting. Such coincidences, if confirmed with high resolution spectra, are usually considered as the signature of line-driven acceleration \citep[][]{Srianand2002}.
The \civ\ absorption from BAL-A shows a large variability during our monitoring. In addition to \civ, we also detect \nv\ and \siiv\
absorption from this component (when \civ\ and \nv\ equivalent widths are large enough). Interestingly \nv\ and \siiv\ vary in unison with \civ.
We find the absorbing gas to cover substantial part of both the continuum and BLR regions implying that the absorbing gas is either co-spatial or located outside the BLR. 
Observed equivalent width variations can be reasonably well reproduced by optical depth and/or covering factor variations. 

Constraining the location of the absorbing gas is important to understand their origin.
Photo-ionization models suggest the ionization parameter to be in the range, 
$\log~U\ge-1.0$. Based on the covering factor and recombination time-scale arguments we suggest the location of the absorbing gas to be $\le113$ pc and outside the BLR (i.e $>0.1$ pc).  Presence of absorption from fine-structure levels will allow us to constrain the distance accurately \citep[for example,][]{Srianand2001}. 
\citet{arav2018} using density sensitive \siv\ troughs found that at least 50\% of the quasar outflows are situated at distances greater than 100 pc and 12\% at distances greater than 1 kpc. Similar analyses \citep{borguet2013,chamberlain2015,arav2020} have used high-ionization diagnostics to show that the outflows are located at several hundred parsecs rather than sub-parsec scales (0.01-0.1 pc) as assumed in standard disk wind models. It will be interesting to search for the presence of fine-structure absorption lines using high resolution and high SNR spectra of \J13\ to constrain the locations of different components with respect to the continuum source.

BAL-B, at velocities of 10500-13300 \kms, is a relatively weaker. 
There are indications that even this component is made up of narrow absorption components. The \siiv\ absorption is not detected from this component and \nv\ and \lya\ are difficult to detect as they are  weak and probably contaminated by intervening  absorptions.

We report the emergence of BAL-C spread over a large velocity range (15000-29000 \kms). Starting from the initial SALT epochs, its rest equivalent width reaches maximum during epoch-6. The \civ\ absorption progressively became weaker in subsequent epochs. The large ejection velocities coupled with large velocity spread suggest a scenario where the emerged gas is launched close to the central engine and may be associated with an accretion disk wind. A few narrow absorption features are also detected in BAL-C at certain epochs indicating the presence of rapidly evolving clumps of gas embedded in a smooth flow. 
We find the observed absorption profiles are inconsistent with  
simple models using density and velocity profiles predicted in typical
1D wind simulations together with photo-ionization calculations.

We note all the identified BAL components show correlated variability. We find a strong anti-correlation between the monochromatic flux at 1640 \AA (F$_{1640 \text{\AA}}$) and the \civ\ rest equivalent width with a large scatter for all the components. We also find the fractional variation in \civ\ equivalent width between any two epochs is correlated with the fractional change in F$_{1640 \text{\AA}}$ between the same epochs. All these are consistent with photo-ionization induced changes to be one of the main causes of the observed line-variability. We also notice similar changes in the strength of the \civ\ BEL which confirms the presence of variations in the ionizing radiation.

We obtained publicly available photometric light curves to look for possible long time-scale variations in the continuum flux which may lead to the photoionization induced BAL variability. Even though we see considerable fading of \J13\ in g, r and i bands during the emergence of BAL-C, the light curve does not show large coherent variation as suggested by photoionization models. 
The photoionization scenario could be reconciled with observations if (i) variation of the \civ\ ionizing photons are larger than that of the UV continuum seen in our spectra and (ii) there is a scatter in the ionizing photon flux for a given UV flux observed.
Moreover, it is also possible that  variations in other parameters such as the covering factor  or changes in the gas properties 
could contribute to the scatter in the measured rest equivalent width for a given measured flux. 

The physical picture we suggest consists of BAL-A being a stable clumpy outflow located far from the central source (further than broad emission line region) and BAL-C being a newly formed wind component
located near the accretion disk.
While it is possible that density profile along our line of sight in the case of BAL-C may be time variable and the covering factor of BAL-A may vary with time, any large variation in the ionizing flux will introduce correlated variability. Continued monitoring (in particular at higher spectral resolution) of this interesting source will provide further insights into this interesting system.

\section*{Acknowledgements}
We thank the anonymous referee for useful suggestions.
We thank  Nishant Singh,  Aseem Paranjape and K. Subramanian for useful discussions. PA thanks Labanya K Guha for helpful discussions on several python programming techniques used in this paper.
PPJ thanks Camille No\^us (Laboratoire Cogitamus) for 
inappreciable and often unnoticed discussions, advice and support.

This paper makes use of SDSS observational data. Funding for the Sloan Digital Sky Survey IV has been provided by the Alfred P. Sloan Foundation, the U.S. Department of Energy Office of Science, and the Participating Institutions. SDSS-IV acknowledges support and resources from the Center for High Performance Computing  at the University of Utah. The SDSS website is www.sdss.org. SDSS-IV is managed by the Astrophysical Research Consortium for the Participating Institutions of the SDSS Collaboration including  the Brazilian Participation Group, the Carnegie Institution for Science, Carnegie Mellon University, Center for Astrophysics | Harvard \& Smithsonian, the Chilean Participation 
Group, the French Participation Group, Instituto de Astrof\'isica de 
Canarias, The Johns Hopkins University, Kavli Institute for the 
Physics and Mathematics of the Universe (IPMU) / University of Tokyo, the Korean Participation Group, Lawrence Berkeley National Laboratory, Leibniz Institut f\"ur Astrophysik Potsdam (AIP),  Max-Planck-Institut 
f\"ur Astronomie (MPIA Heidelberg), Max-Planck-Institut f\"ur Astrophysik (MPA Garching), Max-Planck-Institut f\"ur Extraterrestrische Physik (MPE), National Astronomical Observatories of China, New Mexico State University, 
New York University, University of Notre Dame, Observat\'ario Nacional / MCTI, The Ohio State University, Pennsylvania State University, Shanghai 
Astronomical Observatory, United Kingdom Participation Group, 
Universidad Nacional Aut\'onoma de M\'exico, University of Arizona, University of Colorado Boulder, University of Oxford, University of Portsmouth, University of Utah, University of Virginia, University of Washington, University of 
Wisconsin, Vanderbilt University, and Yale University.

\section*{Data Availability}
Data used in this work are obtained using SALT. Raw data will become available for public use 1.5 years after the observing date at https://ssda.saao.ac.za/.



\bibliographystyle{mnras}
\bibliography{mybib_bal} 





\appendix

\bsp	
\label{lastpage}
\end{document}